\documentstyle[sprocl,epsf]{article}

\begin{document}

%%%%%%%%%%%%%%%%%%%%%%%%%%%%%%%%%%%%%%%%%%%%%%%%%%%%%%%%%
\thispagestyle{empty}

{\large
\begin{flushright}
CLNS 00/1660\\
January 2000\\[0.2cm]
{\tt hep-ph/0001334}
\end{flushright}

\begin{center}
\vspace{1.8cm}
{\huge\bf Introduction to B Physics}\\
\vspace{1.2cm}
Matthias Neubert\\[0.2cm]
{\sl Newman Laboratory of Nuclear Studies, Cornell University\\
Ithaca, New York 14853, USA}\\
\vspace{2.0cm}
{\bf Abstract:}\\[0.2cm]
\parbox{12cm}{These lectures provide an introduction to various topics in 
heavy-flavor physics. We review the theory and phenomenology of 
heavy-quark symmetry, exclusive weak decays of $B$ mesons, inclusive 
decay rates, and some rare $B$ decays.}
\vfill
{\sl Lectures presented at the Trieste Summer School\\ 
in Particle Physics (Part II)\\
Trieste, Italy, 21 June -- 9 July, 1999}
\end{center}}

\newpage
\setcounter{page}{1}
%%%%%%%%%%%%%%%%%%%%%%%%%%%%%%%%%%%%%%%%%%%%%%%%%%%%%%%%%
\title{INTRODUCTION TO B PHYSICS}

\author{MATTHIAS NEUBERT}

\address{Newman Laboratory of Nuclear Studies\\
Cornell University, Ithaca, NY 14853, USA}

\maketitle\abstracts{
These lectures provide an introduction to various topics in 
heavy-flavor physics. We review the theory and phenomenology of 
heavy-quark symmetry, exclusive weak decays of $B$ mesons, inclusive 
decay rates, and some rare $B$ decays.}

\section{Introduction}

The rich phenomenology of weak decays has always been a source of
information about the nature of elementary particle interactions. A
long time ago, $\beta$- and $\mu$-decay experiments revealed the
structure of the effective flavor-changing interactions at low
momentum transfer. Today, weak decays of hadrons containing heavy
quarks are employed for tests of the Standard Model and measurements
of its parameters. In particular, they offer the most direct way to
determine the weak mixing angles, to test the unitarity of the
Cabibbo-Kobayashi-Maskawa (CKM) matrix, and to explore the physics
of CP violation. Hopefully, this will provide some hints about New
Physics beyond the Standard Model. On the other hand, hadronic weak 
decays also serve as a probe of that part of strong-interaction 
phenomenology which is least understood: the confinement of quarks 
and gluons inside hadrons.

The structure of weak interactions in the Standard Model is rather
simple. Flavor-changing decays are mediated by the coupling of the
charged current $J_{\rm CC}^\mu$ to the $W$-boson field:
\begin{equation}
   {\cal L}_{\rm CC} = - {g\over\sqrt{2}}\,J_{\rm CC}^\mu\,
   W_\mu^\dagger + \mbox{h.c.,}
\end{equation}
where
\begin{equation}
   J_{\rm CC}^\mu =
   (\bar\nu_e, \bar\nu_\mu, \bar\nu_\tau)\,\gamma^\mu
   \left( \begin{array}{c} e_{\rm L} \\ \mu_{\rm L} \\ \tau_{\rm L}
   \end{array} \right)
   + (\bar u_{\rm L}, \bar c_{\rm L}, \bar t_{\rm L})\,\gamma^\mu\,
   V_{\rm CKM} \left( \begin{array}{c} d_{\rm L} \\ s_{\rm L} \\
   b_{\rm L} \end{array} \right)
\end{equation}
contains the left-handed lepton and quark fields, and
\begin{equation}
   V_{\rm CKM} = \left( \begin{array}{ccc}
    V_{ud} & V_{us} & V_{ub} \\
    V_{cd} & V_{cs} & V_{cb} \\
    V_{td} & V_{ts} & V_{tb}
   \end{array} \right)
\end{equation}
is the CKM matrix. At low energies, the charged-current interaction
gives rise to local four-fermion couplings of the form
\begin{equation}\label{LFermi}
   {\cal L}_{\rm eff} = - 2\sqrt{2} G_F\,J_{\rm CC}^\mu
   J_{{\rm CC},\mu}^\dagger \,,
\end{equation}
where
\begin{equation}
   G_F = {g^2\over 4\sqrt{2} M_W^2} = 1.16639(2)~\mbox{GeV}^{-2}
\end{equation}
is the Fermi constant.

\begin{figure}[htb]
   \epsfxsize=6cm
   \centerline{\epsffile{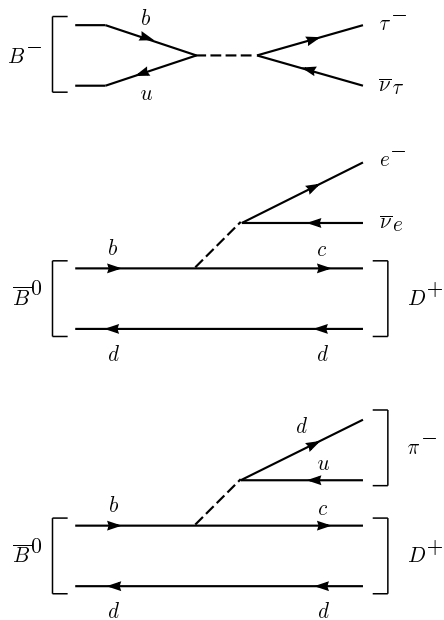}}
\caption{\label{fig:classes}
Examples of leptonic ($B^-\to\tau^-\bar\nu_\tau$), semi-leptonic
($\bar B^0\to D^+ e^-\bar\nu_e$), and non-leptonic ($\bar B^0\to
D^+\pi^-$) decays of $B$ mesons.}
\end{figure}

\begin{figure}[htb]
   \epsfxsize=9.5cm
   \centerline{\epsffile{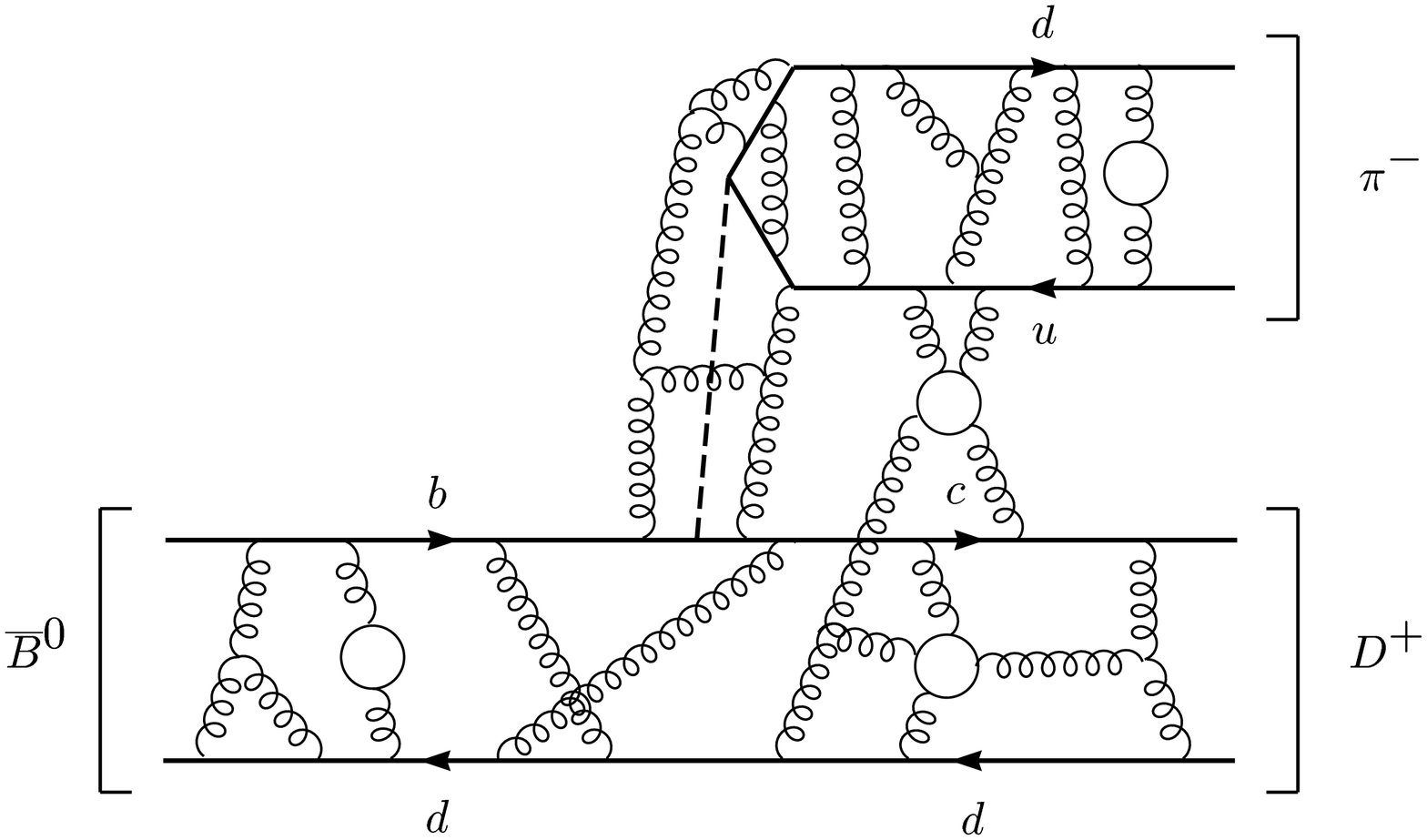}}
\caption{\label{fig:nonlep}
More realistic representation of a non-leptonic decay.}
\end{figure}

According to the structure of the charged-current interaction, weak
decays of had\-rons can be divided into three classes: leptonic
decays, in which the quarks of the decaying hadron annihilate each
other and only leptons appear in the final state; semi-leptonic
decays, in which both leptons and hadrons appear in the final state;
and non-leptonic decays, in which the final state consists of hadrons
only. Representative examples of these three types of decays are
shown in Fig.~\ref{fig:classes}. The simple quark-line graphs shown
in this figure are a gross oversimplification, however. In the real
world, quarks are confined inside hadrons, bound by the exchange of
soft gluons. The simplicity of the weak interactions is
over\-sha\-dowed by the complexity of the strong interactions. A
complicated interplay between the weak and strong forces
characterizes the phenomenology of hadronic weak decays. As an
example, a more realistic picture of a non-leptonic decay is shown in
Fig.~\ref{fig:nonlep}.

The complexity of strong-interaction effects increases with the
number of quarks appearing in the final state. Bound-state effects 
in leptonic decays can be lumped into a single parameter (a ``decay
constant''), while those in semi-leptonic decays are described by
invariant form factors depending on the momentum transfer $q^2$
between the hadrons. Approximate symmetries of the strong
interactions help us to constrain the properties of these form
factors. Non-leptonic weak decays, on the other hand, are much more
complicated to deal with theoretically. Only very recently reliable
tools have been developed that allow us to control the complex QCD 
dynamics in many two-body $B$ decays using a heavy-quark expansion.

Over the last decade, a lot of information on heavy-quark decays has
been collected in experiments at $e^+ e^-$ storage rings operating at
the $\Upsilon(4s)$ resonance, and more recently at high-energy $e^+
e^-$ and hadron colliders. This has led to a rather detailed
knowledge of the flavor sector of the Standard Model and many of the
parameters associated with it. In the years ahead the $B$ factories 
at SLAC, KEK, Cornell, and DESY will continue to provide a wealth of 
new results, focusing primarily on studies of CP violation and rare 
decays.

The experimental progress in heavy-flavor physics has been 
accompanied by a significant progress in theory, which was related to
the discovery of heavy-quark symmetry, the development of the
heavy-quark effective theory, and more generally the establishment of 
various kinds of heavy-quark expansions. The excitement about these 
developments rests upon the fact that they allow model-independent 
predictions in an area in which ``progress'' in theory often meant 
nothing more than the construction of a new model, which could be 
used to estimate some strong-interaction hadronic matrix elements. In 
Sec.~\ref{sec:2}, we review the physical picture behind heavy-quark 
symmetry and discuss the construction, as well as simple applications, 
of the heavy-quark effective theory. Section~\ref{sec:3} deals with
applications of these concepts to exclusive weak decays of $B$
mesons. Applications of the heavy-quark expansion to inclusive $B$ 
decays are reviewed in Sec.~\ref{sec:4}. We then focus on the 
exciting field of rare hadronic $B$ decays, concentrating on the
example of the decays $B\to\pi K$. In Sec.~\ref{sec:5}, we discuss 
the theoretical description of these decays and explain various
strategies for constraining and determining the weak, CP-violating 
phase $\gamma=\mbox{arg}(V_{ub}^*)$ of the CKM matrix. In 
Sec.~\ref{sec:6}, we discuss how rare decays can be used to search 
for New Physics beyond the Standard Model.

\section{Heavy-Quark Symmetry}
\label{sec:2}

This section provides an introduction to the ideas of heavy-quark
symmetry~\cite{Shu1}$^-$\cite{Isgu} and the heavy-quark effective
theory~\cite{EiFe}$^-$\cite{Mann}, which provide the modern
theoretical framework for the description of the properties and
decays of hadrons containing a heavy quark. For a more detailed
description of this subject, the reader is referred to the review
articles in Refs.~18--24.
%\citelow{review}--\citelow{AFrev}.

\subsection{The Physical Picture}

There are several reasons why the strong interactions of hadrons
containing heavy quarks are easier to understand than those of
hadrons containing only light quarks. The first is asymptotic
freedom, the fact that the effective coupling constant of QCD becomes
weak in processes with a large momentum transfer, corresponding to
interactions at short distance scales~\cite{Gros,Poli}. At large
distances, on the other hand, the coupling becomes strong, leading to
non-perturbative phenomena such as the confinement of quarks and
gluons on a length scale $R_{\rm had}\sim 1/\Lambda_{\rm QCD}\sim
1$~fm, which determines the size of hadrons. Roughly speaking,
$\Lambda_{\rm QCD}\sim 0.2$ GeV is the energy scale that separates
the regions of large and small coupling constant. When the mass of a
quark $Q$ is much larger than this scale, $m_Q\gg\Lambda_{\rm QCD}$,
it is called a heavy quark. The quarks of the Standard Model fall
naturally into two classes: up, down and strange are light quarks,
whereas charm, bottom and top are heavy quarks.\footnote{Ironically,
the top quark is of no relevance to our discussion here, since it is
too heavy to form hadronic bound states before it decays.}
For heavy quarks, the effective coupling constant $\alpha_s(m_Q)$ is
small, implying that on length scales comparable to the Compton
wavelength $\lambda_Q\sim 1/m_Q$ the strong interactions are
perturbative and much like the electromagnetic interactions. In fact,
the quarkonium systems $(\bar QQ)$, whose size is of order
$\lambda_Q/\alpha_s(m_Q)\ll R_{\rm had}$, are very much
hydrogen-like.

Systems composed of a heavy quark and other light constituents are
more complicated. The size of such systems is determined by $R_{\rm
had}$, and the typical momenta exchanged between the heavy and light
constituents are of order $\Lambda_{\rm QCD}$. The heavy quark is
surrounded by a complicated, strongly interacting cloud of light
quarks, antiquarks and gluons. In this case it is the fact that
$\lambda_Q\ll R_{\rm had}$, i.e.\ that the Compton wavelength of the
heavy quark is much smaller than the size of the hadron, which leads
to simplifications. To resolve the quantum numbers of the heavy quark
would require a hard probe; the soft gluons exchanged between the
heavy quark and the light constituents can only resolve distances
much larger than $\lambda_Q$. Therefore, the light degrees of freedom
are blind to the flavor (mass) and spin orientation of the heavy
quark. They experience only its color field, which extends over
large distances because of confinement. In the rest frame of the
heavy quark, it is in fact only the electric color field that is
important; relativistic effects such as color magnetism vanish as
$m_Q\to\infty$. Since the heavy-quark spin participates in
interactions only through such relativistic effects, it decouples.

It follows that, in the limit $m_Q\to\infty$, hadronic systems which
differ only in the flavor or spin quantum numbers of the heavy quark
have the same configuration of their light degrees of
freedom~\cite{Shu1}$^-$\cite{Isgu}. Although this observation still
does not allow us to calculate what this configuration is, it
provides relations between the properties of such particles as the
heavy mesons $B$, $D$, $B^*$ and $D^*$, or the heavy baryons
$\Lambda_b$ and $\Lambda_c$ (to the extent that corrections to the
infinite quark-mass limit are small in these systems). These
relations result from some approximate symmetries of the effective
strong interactions of heavy quarks at low energies. The
configuration of light degrees of freedom in a hadron containing a
single heavy quark with velocity $v$ does not change if this quark is
replaced by another heavy quark with different flavor or spin, but
with the same velocity. Both heavy quarks lead to the same static
color field. For $N_h$ heavy-quark flavors, there is thus an SU$(2
N_h)$ spin-flavor symmetry group, under which the effective strong
interactions are invariant. These symmetries are in close
correspondence to familiar properties of atoms. The flavor symmetry
is analogous to the fact that different isotopes have the same
chemistry, since to good approximation the wave function of the
electrons is independent of the mass of the nucleus. The electrons
only see the total nuclear charge. The spin symmetry is analogous to
the fact that the hyperfine levels in atoms are nearly degenerate.
The nuclear spin decouples in the limit $m_e/m_N\to 0$.

Heavy-quark symmetry is an approximate symmetry, and corrections
arise since the quark masses are not infinite. In many respects, it
is complementary to chiral symmetry, which arises in the opposite
limit of small quark masses. There is an important distinction,
however. Whereas chiral symmetry is a symmetry of the QCD Lagrangian
in the limit of vanishing quark masses, heavy-quark symmetry is not a
symmetry of the Lagrangian (not even an approximate one), but rather
a symmetry of an effective theory that is a good approximation to
QCD in a certain kinematic region. It is realized only in systems in
which a heavy quark interacts predominantly by the exchange of soft
gluons. In such systems the heavy quark is almost on-shell; its
momentum fluctuates around the mass shell by an amount of order
$\Lambda_{\rm QCD}$. The corresponding fluctuations in the velocity
of the heavy quark vanish as $\Lambda_{\rm QCD}/m_Q\to 0$. The
velocity becomes a conserved quantity and is no longer a dynamical
degree of freedom~\cite{Geor}. Nevertheless, results derived on the
basis of heavy-quark symmetry are model-independent consequences of
QCD in a well-defined limit. The symmetry-breaking corrections can be
studied in a systematic way. To this end, it is however necessary to
cast the QCD Lagrangian for a heavy quark,
\begin{equation}\label{QCDLag}
   {\cal L}_Q = \bar Q\,(i\rlap{\,/}D - m_Q)\,Q \,,
\end{equation}
into a form suitable for taking the limit $m_Q\to\infty$.

\subsection{Heavy-Quark Effective Theory}

The effects of a very heavy particle often become irrelevant at low
energies. It is then useful to construct a low-energy effective
theory, in which this heavy particle no longer appears. Eventually,
this effective theory will be easier to deal with than the full
theory. A familiar example is Fermi's theory of the weak
interactions. For the description of the weak decays of hadrons, the
weak interactions can be approximated by point-like four-fermion
couplings, governed by a dimensionful coupling constant $G_F$
[cf.~(\ref{LFermi})]. The effects of the intermediate $W$ bosons can 
only be resolved at energies much larger than the hadron masses.

The process of removing the degrees of freedom of a heavy particle
involves the following steps~\cite{SVZ1}$^-$\cite{Polc}: one first
identifies the heavy-particle fields and ``integrates them out'' in
the generating functional of the Green functions of the theory. This
is possible since at low energies the heavy particle does not appear
as an external state. However, whereas the action of the full theory
is usually a local one, what results after this first step is a
non-local effective action. The non-locality is related to the fact
that in the full theory the heavy particle with mass $M$ can appear
in virtual processes and propagate over a short but finite distance
$\Delta x\sim 1/M$. Thus, a second step is required to obtain a local
effective Lagrangian: the non-local effective action is rewritten as
an infinite series of local terms in an Operator Product Expansion
(OPE)~\cite{Wils,Zimm}. Roughly speaking, this corresponds to an
expansion in powers of $1/M$. It is in this step that the short- and
long-distance physics is disentangled. The long-distance physics
corresponds to interactions at low energies and is the same in the
full and the effective theory. But short-distance effects arising
from quantum corrections involving large virtual momenta (of order
$M$) are not described correctly in the effective theory once the 
heavy particle has been integrated out. In a third step, they have to 
be added in a perturbative way using renormalization-group techniques.
These short-distance effects lead to a renormalization of the
coefficients of the local operators in the effective Lagrangian. An
example is the effective Lagrangian for non-leptonic weak decays, in
which radiative corrections from hard gluons with virtual momenta in
the range between $m_W$ and some low renormalization scale $\mu$ 
give rise to Wilson coefficients, which renormalize the local
four-fermion interactions~\cite{AltM}$^-$\cite{Gilm}.

\begin{figure}[htb]
   \epsfysize=7cm
   \centerline{\epsffile{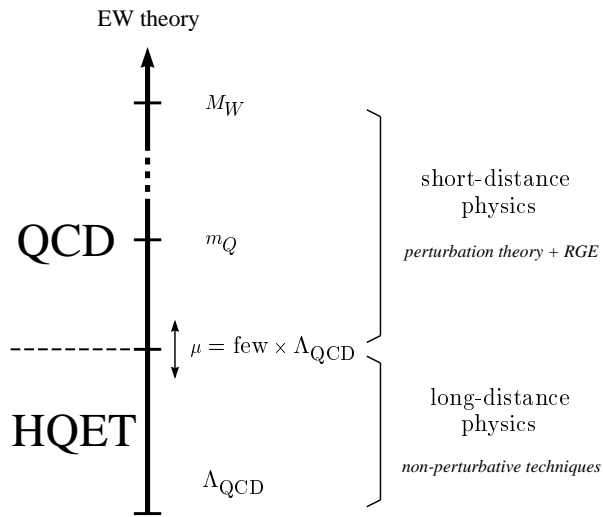}}
\caption{\label{fig:magic}
Philosophy of the heavy-quark effective theory.}
\end{figure}

The heavy-quark effective theory (HQET) is constructed to provide a
simplified description of processes where a heavy quark interacts
with light degrees of freedom predominantly by the exchange of soft
gluons~\cite{EiFe}$^-$\cite{Mann}. Clearly, $m_Q$ is the high-energy
scale in this case, and $\Lambda_{\rm QCD}$ is the scale of the
hadronic physics we are interested in. The situation is illustrated
in Fig.~\ref{fig:magic}. At short distances, i.e.\ for energy scales
larger than the heavy-quark mass, the physics is perturbative and
described by conventional QCD. For mass scales much below the
heavy-quark mass, the physics is complicated and non-perturbative
because of confinement. Our goal is to obtain a simplified
description in this region using an effective field theory. To
separate short- and long-distance effects, we introduce a separation
scale $\mu$ such that $\Lambda_{\rm QCD}\ll\mu\ll m_Q$. The HQET will
be constructed in such a way that it is equivalent to QCD in the
long-distance region, i.e.\ for scales below $\mu$. In the
short-distance region, the effective theory is incomplete, since some
high-momentum modes have been integrated out from the full theory.
The fact that the physics must be independent of the arbitrary scale
$\mu$ allows us to derive renormalization-group equations, which can
be employed to deal with the short-distance effects in an efficient
way.

Compared with most effective theories, in which the degrees of
freedom of a heavy particle are removed completely from the
low-energy theory, the HQET is special in that its purpose is to
describe the properties and decays of hadrons which do contain a
heavy quark. Hence, it is not possible to remove the heavy quark
completely from the effective theory. What is possible is to
integrate out the ``small components'' in the full heavy-quark
spinor, which describe the fluctuations around the mass shell.

The starting point in the construction of the HQET is the observation
that a heavy quark bound inside a hadron moves more or less with the
hadron's velocity $v$ and is almost on-shell. Its momentum can be
written as
\begin{equation}\label{kresdef}
   p_Q^\mu = m_Q v^\mu + k^\mu \,,
\end{equation}
where the components of the so-called residual momentum $k$ are much
smaller than $m_Q$. Note that $v$ is a four-velocity, so that
$v^2=1$. Interactions of the heavy quark with light degrees of
freedom change the residual momentum by an amount of order $\Delta
k\sim\Lambda_{\rm QCD}$, but the corresponding changes in the
heavy-quark velocity vanish as $\Lambda_{\rm QCD}/m_Q\to 0$. In this
situation, it is appropriate to introduce large- and small-component
fields, $h_v$ and $H_v$, by
\begin{equation}\label{hvHvdef}
   h_v(x) = e^{i m_Q v\cdot x}\,P_+\,Q(x) \,, \qquad
   H_v(x) = e^{i m_Q v\cdot x}\,P_-\,Q(x) \,,
\end{equation}
where $P_+$ and $P_-$ are projection operators defined as
\begin{equation}
   P_\pm = {1\pm\rlap/v\over 2} \,.
\end{equation}
It follows that
\begin{equation}\label{redef}
   Q(x) = e^{-i m_Q v\cdot x}\,[ h_v(x) + H_v(x) ] \,.
\end{equation}
Because of the projection operators, the new fields satisfy
$\rlap/v\,h_v=h_v$ and $\rlap/v\,H_v=-H_v$. In the rest frame, i.e.\
for $v^\mu=(1,0,0,0)$, $h_v$ corresponds to the upper two components
of $Q$, while $H_v$ corresponds to the lower ones. Whereas $h_v$
annihilates a heavy quark with velocity $v$, $H_v$ creates a heavy
antiquark with velocity $v$.

\begin{figure}[htb]
   \epsfxsize=7cm
   \centerline{\epsffile{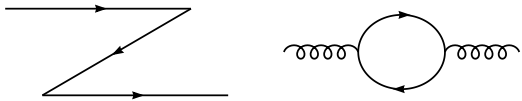}}
\caption{\label{fig:3.1}
Virtual fluctuations involving pair creation of heavy quarks. Time
flows to the right.}
\end{figure}

In terms of the new fields, the QCD Lagrangian (\ref{QCDLag}) for a
heavy quark takes the form
\begin{equation}\label{Lhchi}
   {\cal L}_Q = \bar h_v\,i v\!\cdot\!D\,h_v
   - \bar H_v\,(i v\!\cdot\!D + 2 m_Q)\,H_v
   + \bar h_v\,i\rlap{\,/}D_\perp H_v
   + \bar H_v\,i\rlap{\,/}D_\perp h_v \,,
\end{equation}
where $D_\perp^\mu = D^\mu - v^\mu\,v\cdot D$ is orthogonal to the
heavy-quark velocity: $v\cdot D_\perp=0$. In the rest frame,
$D_\perp^\mu=(0,\vec D\,)$ contains the spatial components of the
covariant derivative. From (\ref{Lhchi}), it is apparent that $h_v$
describes massless degrees of freedom, whereas $H_v$ corresponds to
fluctuations with twice the heavy-quark mass. These are the heavy
degrees of freedom that will be eliminated in the construction of the
effective theory. The fields are mixed by the presence of the third
and fourth terms, which describe pair creation or annihilation of
heavy quarks and antiquarks. As shown in the first diagram in
Fig.~\ref{fig:3.1}, in a virtual process, a heavy quark propagating
forward in time can turn into an antiquark propagating backward in
time, and then turn back into a quark. The energy of the intermediate
quantum state $h h\bar H$ is larger than the energy of the incoming
heavy quark by at least $2 m_Q$. Because of this large energy gap,
the virtual quantum fluctuation can only propagate over a short
distance $\Delta x\sim 1/m_Q$. On hadronic scales set by $R_{\rm
had}=1/\Lambda_{\rm QCD}$, the process essentially looks like a local
interaction of the form
\begin{equation}
   \bar h_v\,i\rlap{\,/}D_\perp\,{1\over 2 m_Q}\,
   i\rlap{\,/}D_\perp h_v \,,
\end{equation}
where we have simply replaced the propagator for $H_v$ by $1/2 m_Q$.
A more correct treatment is to integrate out the small-component
field $H_v$, thereby deriving a non-local effective action for the
large-component field $h_v$, which can then be expanded in terms of
local operators. Before doing this, let us mention a second type of
virtual corrections involving pair creation, namely heavy-quark
loops. An example is shown in the second diagram in 
Fig.~\ref{fig:3.1}. Heavy-quark loops cannot be described in terms of
the effective fields $h_v$ and $H_v$, since the quark velocities
inside a loop are not conserved and are in no way related to hadron
velocities. However, such short-distance processes are proportional
to the small coupling constant $\alpha_s(m_Q)$ and can be calculated
in perturbation theory. They lead to corrections that are added onto
the low-energy effective theory in the renormalization procedure.

On a classical level, the heavy degrees of freedom represented by
$H_v$ can be eliminated using the equation of motion. Taking the
variation of the Lagrangian with respect to the field $\bar H_v$, we
obtain
\begin{equation}
   (i v\!\cdot\!D + 2 m_Q)\,H_v = i\rlap{\,/}D_\perp h_v \,.
\end{equation}
This equation can formally be solved to give
\begin{equation}\label{Hfield}
   H_v = {1\over 2 m_Q + i v\!\cdot\!D}\,
   i\rlap{\,/}D_\perp h_v \,,
\end{equation}
showing that the small-component field $H_v$ is indeed of order
$1/m_Q$. We can now insert this solution into (\ref{Lhchi}) to obtain
the ``non-local effective Lagrangian''
\begin{equation}\label{Lnonloc}
   {\cal L}_{\rm eff} = \bar h_v\,i v\!\cdot\!D\,h_v
   + \bar h_v\,i\rlap{\,/}D_\perp\,{1\over 2 m_Q+i v\!\cdot\!D}\,
   i\rlap{\,/}D_\perp h_v \,.
\end{equation}
Clearly, the second term corresponds to the first class of virtual
processes shown in Fig.~\ref{fig:3.1}.

It is possible to derive this Lagrangian in a more elegant way by
manipulating the generating functional for QCD Green functions
containing heavy-quark fields~\cite{Mann}. To this end, one starts
from the field redefinition (\ref{redef}) and couples the
large-component fields $h_v$ to external sources $\rho_v$. Green
functions with an arbitrary number of $h_v$ fields can be constructed
by taking derivatives with respect to $\rho_v$. No sources are needed
for the heavy degrees of freedom represented by $H_v$. The functional
integral over these fields is Gaussian and can be performed
explicitly, leading to the effective action
\begin{equation}\label{SeffMRR}
   S_{\rm eff} = \int\!{\rm d}^4 x\,{\cal L}_{\rm eff}
   - i \ln\Delta \,,
\end{equation}
with ${\cal L}_{\rm eff}$ as given in (\ref{Lnonloc}). The
appearance
of the logarithm of the determinant
\begin{equation}
   \Delta = \exp\bigg( {1\over 2}\,{\rm Tr}\,
   \ln\big[ 2 m_Q + i v\!\cdot\!D - i\eta \big] \bigg)
\end{equation}
is a quantum effect not present in the classical derivation presented
above. However, in this case the determinant can be regulated in a
gauge-invariant way, and by choosing the gauge $v\cdot A=0$ one can
show that $\ln\Delta$ is just an irrelevant
constant~\cite{Mann,Soto}.

\begin{figure}[htb]
   \epsfysize=3.5cm
   \centerline{\epsffile{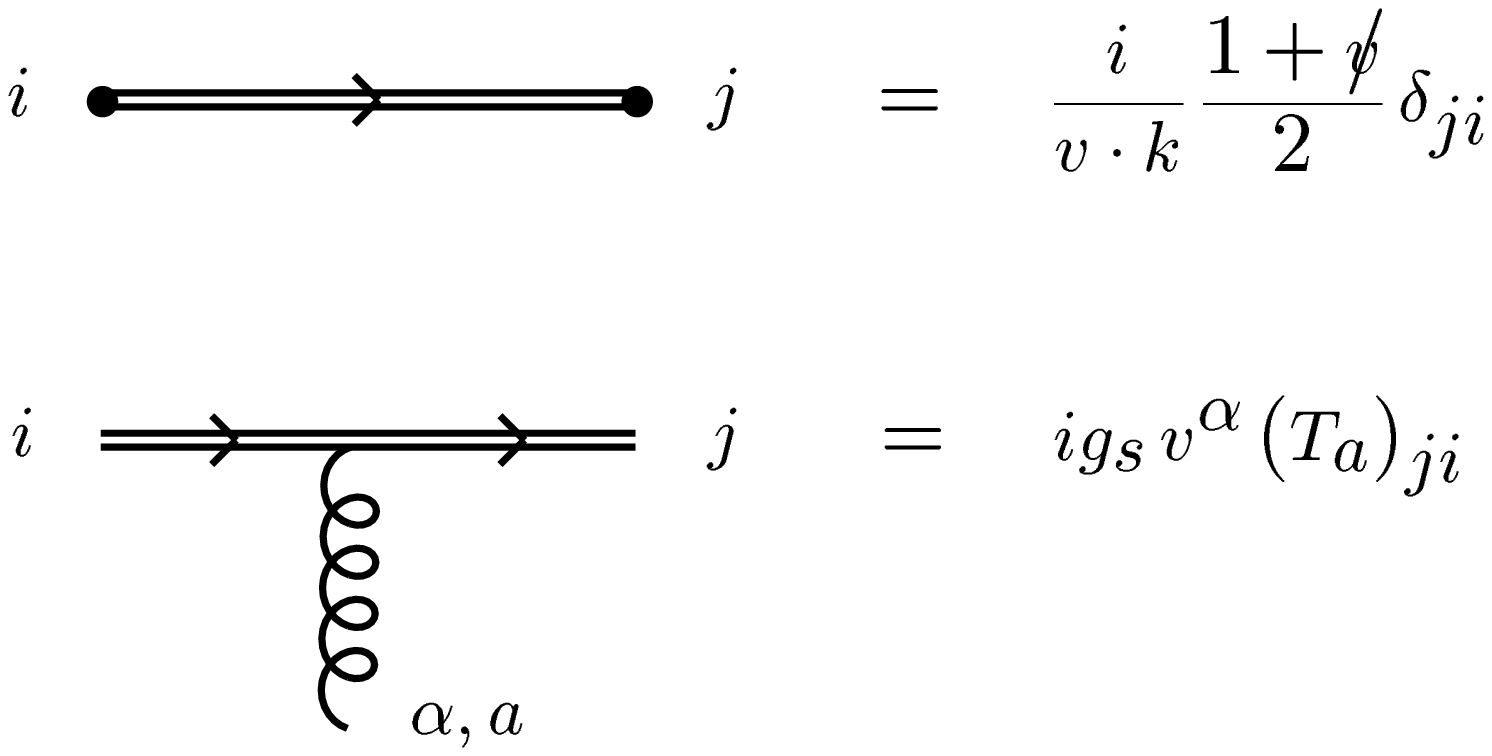}}
\caption{\label{fig:3.2}
Feynman rules of the HQET ($i,j$ and $a$ are color indices). A heavy
quark with velocity $v$ is represented by a double line. The residual
momentum $k$ is defined in (\protect\ref{kresdef}).}
\end{figure}

Because of the phase factor in (\ref{redef}), the $x$ dependence of
the effective heavy-quark field $h_v$ is weak. In momentum space,
derivatives acting on $h_v$ produce powers of the residual momentum
$k$, which is much smaller than $m_Q$. Hence, the non-local effective
Lagrangian (\ref{Lnonloc}) allows for a derivative expansion:
\begin{equation}
   {\cal L}_{\rm eff} = \bar h_v\,i v\!\cdot\!D\,h_v
   + {1\over 2 m_Q}\,\sum_{n=0}^\infty\,
   \bar h_v\,i\rlap{\,/}D_\perp\,\bigg( -{i v\cdot D\over 2 m_Q}
   \bigg)^n\,i\rlap{\,/}D_\perp h_v \,.
\end{equation}
Taking into account that $h_v$ contains a $P_+$ projection operator,
and using the identity
\begin{equation}\label{pplusid}
   P_+\,i\rlap{\,/}D_\perp\,i\rlap{\,/}D_\perp P_+
   = P_+\,\bigg[ (i D_\perp)^2 + {g_s\over 2}\,
   \sigma_{\mu\nu }\,G^{\mu\nu } \bigg]\,P_+ \,,
\end{equation}
where $i[D^\mu,D^\nu]=g_s\,G^{\mu\nu}$ is the gluon field-strength
tensor, one finds that~\cite{EiH1,FGL}
\begin{equation}\label{Lsubl}
   {\cal L}_{\rm eff} = \bar h_v\,i v\cdot\!D\,h_v
   + {1\over 2 m_Q}\,\bar h_v\,(i D_\perp)^2\,h_v
   + {g_s\over 4 m_Q}\,\bar h_v\,\sigma_{\mu\nu}\,
   G^{\mu\nu}\,h_v + O(1/m_Q^2) \,.
\end{equation}
In the limit $m_Q\to\infty$, only the first term remains:
\begin{equation}\label{Leff}
   {\cal L}_\infty = \bar h_v\,i v\!\cdot\!D\,h_v \,.
\end{equation}
This is the effective Lagrangian of the HQET. It gives rise to the
Feynman rules shown in Fig.~\ref{fig:3.2}.

Let us take a moment to study the symmetries of this
Lagrangian~\cite{Geor}. Since there appear no Dirac matrices,
interactions of the heavy quark with gluons leave its spin unchanged.
Associated with this is an SU(2) symmetry group, under which
${\cal L}_\infty$ is invariant. The action of this symmetry on the
heavy-quark fields becomes most transparent in the rest frame, where
the generators $S^i$ of SU(2) can be chosen as
\begin{equation}\label{Si}
   S^i = {1\over 2} \left( \begin{array}{cc}
                           \sigma^i ~&~ 0 \\
                           0 ~&~ \sigma^i \end{array} \right) \,;
   \qquad [S^i,S^j] = i \epsilon^{ijk} S^k \,.
\end{equation}
Here $\sigma^i$ are the Pauli matrices. An infinitesimal SU(2)
transformation $h_v\to (1 + i\vec\epsilon \cdot\vec S\,)\,h_v$ leaves
the Lagrangian invariant:
\begin{equation}\label{SU2tr}
   \delta{\cal L}_\infty = \bar h_v\,
   [i v\!\cdot\! D,i \vec\epsilon\cdot\vec S\,]\,h_v = 0 \,.
\end{equation}
Another symmetry of the HQET arises since the mass of the heavy quark
does not appear in the effective Lagrangian. For $N_h$ heavy quarks
moving at the same velocity, eq.~(\ref{Leff}) can be extended by
writing
\begin{equation}\label{Leff2}
   {\cal L}_\infty
   = \sum_{i=1}^{N_h} \bar h_v^i\,i v\!\cdot\! D\,h_v^i \,.
\end{equation}
This is invariant under rotations in flavor space. When combined
with the spin symmetry, the symmetry group is promoted to SU$(2
N_h)$. This is the heavy-quark spin-flavor symmetry~\cite{Isgu,Geor}.
Its physical content is that, in the limit $m_Q\to\infty$, the strong
interactions of a heavy quark become independent of its mass and
spin.

Consider now the operators appearing at order $1/m_Q$ in the
effective Lagrangian (\ref{Lsubl}). They are easiest to identify in
the rest frame. The first operator,
\begin{equation}\label{Okin}
   {\cal O}_{\rm kin} = {1\over 2 m_Q}\,\bar h_v\,(i D_\perp)^2\,
   h_v \to - {1\over 2 m_Q}\,\bar h_v\,(i \vec D\,)^2\,h_v \,,
\end{equation}
is the gauge-covariant extension of the kinetic energy arising from
the residual motion of the heavy quark. The second operator is the
non-Abelian analogue of the Pauli interaction, which describes the
color-magnetic coupling of the heavy-quark spin to the gluon field:
\begin{equation}\label{Omag}
   {\cal O}_{\rm mag} = {g_s\over 4 m_Q}\,\bar h_v\,
   \sigma_{\mu\nu}\,G^{\mu\nu}\,h_v \to
   - {g_s\over m_Q}\,\bar h_v\,\vec S\!\cdot\!\vec B_c\,h_v \,.
\end{equation}
Here $\vec S$ is the spin operator defined in (\ref{Si}), and $B_c^i
= -\frac{1}{2}\epsilon^{ijk} G^{jk}$ are the components of the
color-magnetic field. The chromo-magnetic interaction is a
relativistic effect, which scales like $1/m_Q$. This is the origin of
the heavy-quark spin symmetry.

\subsection{The Residual Mass Term and the Definition of the
Heavy-Quark Mass}

The choice of the expansion parameter in the HQET, i.e.\ the
definition of the heavy-quark mass $m_Q$, deserves some comments. In
the derivation presented earlier in this section, we chose $m_Q$ to
be the ``mass in the Lagrangian'', and using this parameter in the
phase redefinition in (\ref{redef}) we obtained the effective
Lagrangian (\ref{Leff}), in which the heavy-quark mass no longer
appears. However, this treatment has its subtleties. The symmetries
of the HQET allow a ``residual mass'' $\delta m$ for the heavy quark,
provided that $\delta m$ is of order $\Lambda_{\rm QCD}$ and is the
same for all heavy-quark flavors. Even if we arrange that such a
mass term is not present at the tree level, it will in general be
induced by quantum corrections. (This is unavoidable if the theory is
regulated with a dimensionful cutoff.) Therefore, instead of
(\ref{Leff}) we should write the effective Lagrangian in the more
general form~\cite{FNL}
\begin{equation}
   {\cal L}_\infty = \bar h_v\,iv\!\cdot\!D\,h_v
   - \delta m\,\bar h_v h_v \,.
\end{equation}
If we redefine the expansion parameter according to $m_Q\to
m_Q+\Delta m$, the residual mass changes in the opposite way: $\delta
m\to\delta m-\Delta m$. This implies that there is a unique choice of
the expansion parameter $m_Q$ such that $\delta m=0$. Requiring
$\delta m=0$, as it is usually done implicitly in the HQET, defines a
heavy-quark mass, which in perturbation theory coincides with the
pole mass~\cite{Tarr}. This, in turn, defines for each heavy hadron
$H_Q$ a parameter $\bar\Lambda$ (sometimes called the ``binding
energy'') through
\begin{equation}
   \bar\Lambda = (m_{H_Q} - m_Q)\Big|_{m_Q\to\infty} \,.
\label{Lbdef}
\end{equation}
If one prefers to work with another choice of the expansion
parameter, the values of non-perturbative parameters such as
$\bar\Lambda$ change, but at the same time one has to include the
residual mass term in the HQET Lagrangian. It can be shown that the
various parameters depending on the definition of $m_Q$ enter the
predictions for physical quantities in such a way that the results
are independent of the particular choice adopted~\cite{FNL}.

There is one more subtlety hidden in the above discussion. The
quantities $m_Q$, $\bar\Lambda$ and $\delta m$ are non-perturbative
parameters of the HQET, which have a similar status as the vacuum
condensates in QCD phenomenology~\cite{SVZ}. These parameters cannot
be defined unambiguously in perturbation theory. The reason lies in
the divergent behavior of perturbative expansions in large orders,
which is associated with the existence of singularities along the
real axis in the Borel plane, the so-called
renormalons~\cite{tHof}$^-$\cite{Muel}. For instance, the
perturbation series which relates the pole mass $m_Q$ of a heavy
quark to its bare mass,
\begin{equation}
   m_Q = m_Q^{\rm bare}\,\Big\{ 1 + c_1\,\alpha_s(m_Q)
   + c_2\,\alpha_s^2(m_Q) + \dots + c_n\,\alpha_s^n(m_Q)
   + \dots \Big\} \,,
\end{equation}
contains numerical coefficients $c_n$ that grow as $n!$ for large
$n$, rendering the series divergent and not Borel
summable~\cite{BBren,Bigiren}. The best one can achieve is to
truncate the perturbation series at its minimal term, but this leads
to an unavoidable arbitrariness of order $\Delta m_Q\sim\Lambda_{\rm
QCD}$ (the size of the minimal term) in the value of the pole mass.
This observation, which at first sight seems a serious problem for
QCD phenomenology, should not come as a surprise. We know that
because of confinement quarks do not appear as physical states in
nature. Hence, there is no unique way to define their on-shell
properties such as a pole mass. Remarkably, QCD perturbation theory 
``knows'' about its incompleteness and indicates, through the 
appearance of renormalon singularities, the presence of 
non-perturbative effects. One must first specify a scheme how to 
truncate the QCD perturbation series before non-perturbative 
statements such as $\delta m=0$ become meaningful, and hence before 
non-perturbative parameters such as $m_Q$ and $\bar\Lambda$ become 
well-defined quantities. The actual values of these parameters will 
depend on this scheme.

We stress that the ``renormalon ambiguities'' are not a conceptual
problem for the heavy-quark expansion. In fact, it can be shown quite
generally that these ambiguities cancel in all predictions for
physical observables~\cite{Chris}$^-$\cite{LMS}. The way the
cancellations occur is intricate, however. The generic structure of
the heavy-quark expansion for an observable is of the form:
\begin{equation}
   \mbox{Observable} \sim C[\alpha_s(m_Q)]\,\bigg( 1
   + {\Lambda\over m_Q} + \dots \bigg) \,,
\end{equation}
where $C[\alpha_s(m_Q)]$ represents a perturbative coefficient
function, and $\Lambda$ is a dimensionful non-perturbative parameter.
The truncation of the perturbation series defining the coefficient
function leads to an arbitrariness of order $\Lambda_{\rm QCD}/m_Q$,
which cancels against a corresponding arbitrariness of order
$\Lambda_{\rm QCD}$ in the definition of the non-perturbative
parameter $\Lambda$.

The renormalon problem poses itself when one imagines to apply
perturbation theory to very high orders. In practice, the
perturbative coefficients are known to finite order in $\alpha_s$ 
(typically to one- or two-loop accuracy), and to be consistent one 
should use them in connection with the pole mass (and $\bar\Lambda$ 
etc.) defined to the same order.

\subsection{Spectroscopic Implications}

The spin-flavor symmetry leads to many interesting relations between
the properties of hadrons containing a heavy quark. The most direct
consequences concern the spectroscopy of such states~\cite{IsWi,Agli}.
In the limit $m_Q\to\infty$, the spin of the heavy quark and the total
angular momentum $j$ of the light degrees of freedom are separately
conserved by the strong interactions. Because of heavy-quark symmetry,
the dynamics is independent of the spin and mass of the heavy quark.
Hadronic states can thus be classified by the quantum numbers (flavor,
spin, parity, etc.) of their light degrees of freedom~\cite{AFal}. The
spin symmetry predicts that, for fixed $j\neq 0$, there is a doublet of
degenerate states with total spin $J=j\pm\frac{1}{2}$. The flavor
symmetry relates the properties of states with different heavy-quark
flavor.

In general, the mass of a hadron $H_Q$ containing a heavy quark $Q$
obeys an expansion of the form
\begin{equation}\label{massexp}
   m_{H_Q} = m_Q + \bar\Lambda + {\Delta m^2\over 2 m_Q}
   + O(1/m_Q^2) \,.
\end{equation}
The parameter $\bar\Lambda$ represents contributions arising from
terms in the Lagrangian that are independent of the heavy-quark
mass~\cite{FNL}, whereas the quantity $\Delta m^2$ originates from
the terms of order $1/m_Q$ in the effective Lagrangian of the HQET.
For the ground-state pseudoscalar and vector mesons, one can
parame\-trize the contributions from the kinetic energy and the
chromo-magnetic interaction in terms of two quantities $\lambda_1$
and $\lambda_2$, in such a way that~\cite{FaNe}
\begin{equation}\label{FNrela}
   \Delta m^2 = -\lambda_1 + 2 \Big[ J(J+1) - \textstyle{3\over 2}
   \Big]\,\lambda_2 \,.
\end{equation}
The hadronic parameters $\bar\Lambda$, $\lambda_1$ and $\lambda_2$
are independent of $m_Q$. They characterize the properties of the
light constituents.

Consider, as a first example, the SU(3) mass splittings for heavy
mesons. The heavy-quark expansion predicts that
\begin{eqnarray}
   m_{B_S} - m_{B_d} &=& \bar\Lambda_s - \bar\Lambda_d
    + O(1/m_b) \,, \nonumber\\
   m_{D_S} - m_{D_d} &=& \bar\Lambda_s - \bar\Lambda_d
    + O(1/m_c) \,,
\end{eqnarray}
where we have indicated that the value of the parameter $\bar\Lambda$
depends on the flavor of the light quark. Thus, to the extent that
the charm and bottom quarks can both be considered sufficiently
heavy, the mass splittings should be similar in the two systems. This
prediction is confirmed experimentally, since
\begin{eqnarray}
   m_{B_S} - m_{B_d} &=& (90\pm 3)~\mbox{MeV} \,, \nonumber\\
   m_{D_S} - m_{D_d} &=& (99\pm 1)~\mbox{MeV} \,.
\end{eqnarray}

As a second example, consider the spin splittings between the
ground-state pseudoscalar ($J=0$) and vector ($J=1$) mesons, which
are the members of the spin-doublet with $j=\frac{1}{2}$. From
(\ref{massexp}) and (\ref{FNrela}), it follows that
\begin{eqnarray}
   m_{B^*}^2 - m_B^2 &=& 4\lambda_2 + O(1/m_b) \,, \nonumber\\
   m_{D^*}^2 - m_D^2 &=& 4\lambda_2 + O(1/m_c) \,.
\end{eqnarray}
The data are compatible with this:
\begin{eqnarray}\label{VPexp}
   m_{B^*}^2 - m_B^2 &\approx& 0.49~{\rm GeV}^2 \,, \nonumber\\
   m_{D^*}^2 - m_D^2 &\approx& 0.55~{\rm GeV}^2 \,.
\end{eqnarray}
Assuming that the $B$ system is close to the heavy-quark limit, we
obtain the value
\begin{equation}
   \lambda_2\approx 0.12~\mbox{GeV}^2
\label{lam2val}
\end{equation}
for one of the hadronic parameters in (\ref{FNrela}). This quantity
plays an important role in the phenomenology of inclusive decays of
heavy hadrons.

A third example is provided by the mass splittings between the
ground-state mesons and baryons containing a heavy quark. The HQET
predicts that
\begin{eqnarray}\label{barmes}
   m_{\Lambda_b} - m_B &=& \bar\Lambda_{\rm baryon}
    - \bar\Lambda_{\rm meson} + O(1/m_b) \,, \nonumber\\
   m_{\Lambda_c} - m_D &=& \bar\Lambda_{\rm baryon}
    - \bar\Lambda_{\rm meson} + O(1/m_c) \,.
\end{eqnarray}
This is again consistent with the experimental results
\begin{eqnarray}
   m_{\Lambda_b} - m_B &=& (345\pm 9)~\mbox{MeV} \,, \nonumber\\
   m_{\Lambda_c} - m_D &=& (416\pm 1)~\mbox{MeV} \,,
\end{eqnarray}
although in this case the data indicate sizeable symmetry-breaking
corrections. The dominant correction to the relations (\ref{barmes}) 
comes from the contribution of the chromo-magnetic interaction to the 
masses of the heavy mesons,\footnote{Because of spin symmetry, there 
is no such contribution to the masses of $\Lambda_Q$ baryons.} 
which adds a term $3\lambda_2/2 m_Q$ on the right-hand side. Including 
this term, we obtain the refined prediction that the two quantities
\begin{eqnarray}
   m_{\Lambda_b} - m_B - {3\lambda_2\over 2 m_B}
   &=& (311\pm 9)~\mbox{MeV} \,, \nonumber\\
   m_{\Lambda_c} - m_D - {3\lambda_2\over 2 m_D}
   &=& (320\pm 1)~\mbox{MeV}
\end{eqnarray}
should be close to each other. This is clearly satisfied by the data.

The mass formula (\ref{massexp}) can also be used to derive
information on the heavy-quark masses from the observed hadron
masses. Introducing the ``spin-averaged'' meson masses
$\overline{m}_B=\frac{1}{4}\,(m_B+3 m_{B^*})\approx 5.31$~GeV and
$\overline{m}_D=\frac{1}{4}\,(m_D+3 m_{D^*})\approx 1.97$~GeV, we
find that
\begin{equation}\label{mbmc}
   m_b-m_c = (\overline{m}_B-\overline{m}_D)\,\bigg\{
   1 - {\lambda_1\over 2\overline{m}_B\overline{m}_D}
   + O(1/m_Q^3) \bigg\} \,.
\end{equation}
Using theoretical estimates for the parameter $\lambda_1$, which lie
in the range~\cite{ElSh}$^-$\cite{Fazi}
\begin{equation}\label{lam1}
   \lambda_1 = -(0.3\pm 0.2)~\mbox{GeV}^2 \,,
\end{equation}
this relation leads to
\begin{equation}\label{mbmcval}
   m_b - m_c = (3.39\pm 0.03\pm 0.03)~\mbox{GeV} \,,
\label{mQdif}
\end{equation}
where the first error reflects the uncertainty in the value of
$\lambda_1$, and the second one takes into account unknown
higher-order corrections. The fact that the difference of the pole 
masses, $m_b-m_c$, is known rather precisely is important for the 
analysis of inclusive decays of heavy hadrons.

\section{Exclusive Semi-Leptonic Decays}
\label{sec:3}

\begin{figure}[htb]
   \epsfxsize=6.2cm
   \centerline{\epsffile{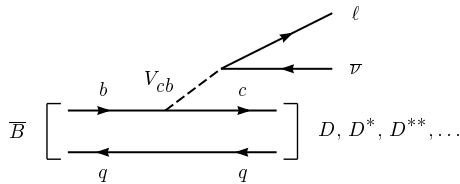}}
\caption{\label{fig:1}
Semi-leptonic decays of $B$ mesons.}
\end{figure}

Semi-leptonic decays of $B$ mesons have received a lot of attention in
recent years. The decay channel $\bar B\to D^*\ell\,\bar\nu$ has the
largest branching fraction of all $B$-meson decay modes. From a
theoretical point of view, semi-leptonic decays are simple enough to
allow for a reliable, quantitative description. The analysis of these
decays provides much information about the strong forces that bind
the quarks and gluons into hadrons. Schematically, a semi-leptonic
decay process is shown in Fig.~\ref{fig:1}. The strength of the $b\to
c$ transition vertex is governed by the element $V_{cb}$ of the CKM
matrix. The parameters of this matrix are fundamental parameters of
the Standard Model. A primary goal of the study of semi-leptonic
decays of $B$ mesons is to extract with high precision the values of
$|V_{cb}|$ and $|V_{ub}|$. We will now discuss the theoretical basis
of such analyses.

\subsection{Weak Decay Form Factors}

Heavy-quark symmetry implies relations between the weak decay form
factors of heavy mesons, which are of particular interest. These
relations have been derived by Isgur and Wise~\cite{Isgu},
generalizing ideas developed by Nussinov and Wetzel~\cite{Nuss}, and
by Voloshin and Shifman~\cite{Vol1,Vol2}.

\begin{figure}[htb]
   \epsfxsize=7cm
   \centerline{\epsffile{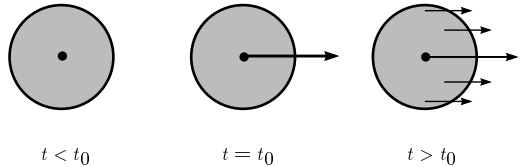}}
\caption{\label{fig:3.3}
Elastic transition induced by an external heavy-quark current.}
\end{figure}

Consider the elastic scattering of a $B$ meson, $\bar B(v)\to\bar
B(v')$, induced by a vector current coupled to the $b$ quark. Before
the action of the current, the light degrees of freedom inside the
$B$ meson orbit around the heavy quark, which acts as a static source
of color. On average, the $b$ quark and the $B$ meson have the same
velocity $v$. The action of the current is to replace instantaneously
(at time $t=t_0$) the color source by one moving at a velocity $v'$,
as indicated in Fig.~\ref{fig:3.3}. If $v=v'$, nothing happens; the
light degrees of freedom do not realize that there was a current
acting on the heavy quark. If the velocities are different, however,
the light constituents suddenly find themselves interacting with a
moving color source. Soft gluons have to be exchanged to rearrange
them so as to form a $B$ meson moving at velocity $v'$. This
rearrangement leads to a form-factor suppression, reflecting the fact
that, as the velocities become more and more different, the
probability for an elastic transition decreases. The important
observation is that, in the limit $m_b\to\infty$, the form factor can
only depend on the Lorentz boost $\gamma = v\cdot v'$ connecting the
rest frames of the initial- and final-state mesons. Thus, in this
limit a dimensionless probability function $\xi(v\cdot v')$ describes
the transition. It is called the Isgur-Wise function~\cite{Isgu}. In
the HQET, which provides the appropriate framework for taking the
limit $m_b\to\infty$, the hadronic matrix element describing the
scattering process can thus be written as
\begin{equation}\label{elast}
   {1\over m_B}\,\langle\bar B(v')|\,\bar b_{v'}\gamma^\mu b_v\,
   |\bar B(v)\rangle = \xi(v\cdot v')\,(v+v')^\mu \,.
\end{equation}
Here $b_v$ and $b_{v'}$ are the velocity-dependent heavy-quark
fields of the HQET. It is important that the function $\xi(v\cdot
v')$ does not depend on $m_b$. The factor $1/m_B$ on the left-hand
side compensates for a trivial dependence on the heavy-meson mass
caused by the relativistic normalization of meson states, which is
conventionally taken to be
\begin{equation}\label{nonrelnorm}
   \langle\bar B(p')|\bar B(p)\rangle = 2 m_B v^0\,(2\pi)^3\,
   \delta^3(\vec p-\vec p\,') \,.
\end{equation}
Note that there is no term proportional to $(v-v')^\mu$ in
(\ref{elast}). This can be seen by contracting the matrix element
with $(v-v')_\mu$, which must give zero since $\rlap/v\,b_v = b_v$
and
$\bar b_{v'}\rlap/v' = \bar b_{v'}$.

It is more conventional to write the above matrix element in terms of
an elastic form factor $F_{\rm el}(q^2)$ depending on the momentum
transfer $q^2=(p-p')^2$:
\begin{equation}
   \langle\bar B(v')|\,\bar b\,\gamma^\mu b\,|\bar B(v)\rangle
   = F_{\rm el}(q^2)\,(p+p')^\mu \,,
\end{equation}
where $p^(\phantom{}'\phantom{}^)=m_B v^(\phantom{}'\phantom{}^)$.
Comparing this with (\ref{elast}), we find that
\begin{equation}
   F_{\rm el}(q^2) = \xi(v\cdot v') \,, \qquad
   q^2 = -2 m_B^2 (v\cdot v'-1) \,.
\end{equation}
Because of current conservation, the elastic form factor is
normalized to unity at $q^2=0$. This condition implies the
normalization of the Isgur-Wise function at the kinematic point
$v\cdot v'=1$, i.e.\ for $v=v'$:
\begin{equation}\label{Jcons2}
   \xi(1) = 1 \,.
\end{equation}
It is in accordance with the intuitive argument that the probability
for an elastic transition is unity if there is no velocity change.
Since for $v=v'$ the final-state meson is at rest in the rest frame
of
the initial meson, the point $v\cdot v'=1$ is referred to as the
zero-recoil limit.

The heavy-quark flavor symmetry can be used to replace the $b$ quark
in the final-state meson by a $c$ quark, thereby turning the $B$
meson into a $D$ meson. Then the scattering process turns into a weak
decay process. In the infinite-mass limit, the replacement $b_{v'}\to
c_{v'}$ is a symmetry transformation, under which the effective
Lagrangian is invariant. Hence, the matrix element
\begin{equation}
   {1\over\sqrt{m_B m_D}}\,\langle D(v')|\,\bar c_{v'}\gamma^\mu
   b_v\,|\bar B(v)\rangle = \xi(v\cdot v')\,(v+v')^\mu
\end{equation}
is still determined by the same function $\xi(v\cdot v')$. This is
interesting, since in general the matrix element of a
flavor-changing current between two pseudoscalar mesons is described
by two form factors:
\begin{equation}
   \langle D(v')|\,\bar c\,\gamma^\mu b\,|\bar B(v)\rangle
   = f_+(q^2)\,(p+p')^\mu - f_-(q^2)\,(p-p')^\mu \,.
\end{equation}
Comparing the above two equations, we find that
\begin{eqnarray}\label{inelast}
   f_\pm(q^2) &=& {m_B\pm m_D\over 2\sqrt{m_B m_D}}\,\xi(v\cdot v')
    \,, \nonumber\\
   q^2 &=& m_B^2 + m_D^2 - 2 m_B m_D\,v\cdot v' \,.
\end{eqnarray}
Thus, the heavy-quark flavor symmetry relates two a priori
independent form factors to one and the same function. Moreover, the
normalization of the Isgur-Wise function at $v\cdot v'=1$ now
implies a non-trivial normalization of the form factors $f_\pm(q^2)$
at the point of maximum momentum transfer, $q_{\rm max}^2=
(m_B-m_D)^2$:
\begin{equation}
   f_\pm(q_{\rm max}^2) = {m_B\pm m_D\over 2\sqrt{m_B m_D}} \,.
\end{equation}

The heavy-quark spin symmetry leads to additional relations among
weak decay form factors. It can be used to relate matrix elements
involving vector mesons to those involving pseudoscalar mesons. A
vector meson with longitudinal polarization is related to a
pseudoscalar meson by a rotation of the heavy-quark spin. Hence, the
spin-symmetry transformation $c_{v'}^\Uparrow\to c_{v'}^\Downarrow$
relates $\bar B\to D$ with $\bar B\to D^*$ transitions. The result of
this transformation is~\cite{Isgu}
\begin{eqnarray}
   {1\over\sqrt{m_B m_{D^*}}}\,
   \langle D^*(v',\varepsilon)|\,\bar c_{v'}\gamma^\mu b_v\,
   |\bar B(v)\rangle &=& i\epsilon^{\mu\nu\alpha\beta}\,
    \varepsilon_\nu^*\,v'_\alpha v_\beta\,\,\xi(v\cdot v') \,,
    \nonumber\\
   {1\over\sqrt{m_B m_{D^*}}}\,
   \langle D^*(v',\varepsilon)|\,\bar c_{v'}\gamma^\mu\gamma_5\,
   b_v\,|\bar B(v)\rangle &=& \Big[ \varepsilon^{*\mu}\,(v\cdot v'+1)
    - v'^\mu\,\varepsilon^*\!\cdot v \Big] \xi(v\cdot v') \,,
    \nonumber\\
\end{eqnarray}
where $\varepsilon$ denotes the polarization vector of the $D^*$
meson. Once again, the matrix elements are completely described in
terms of the Isgur-Wise function. Now this is even more remarkable,
since in general four form factors, $V(q^2)$ for the vector current,
and $A_i(q^2)$, $i=0,1,2$, for the axial current, are required to
parameterize these matrix elements. In the heavy-quark limit, they
obey the relations~\cite{Neu1}
\begin{eqnarray}\label{PVff}
   {m_B+m_{D^*}\over 2\sqrt{m_B m_{D^*}}}\,\xi(v\cdot v')
   &=& V(q^2) = A_0(q^2) = A_1(q^2) \nonumber\\
   &=& \bigg[ 1 - {q^2\over(m_B+m_D)^2} \bigg]^{-1}\,A_1(q^2) \,,
    \nonumber\\
   \phantom{ \Bigg[ }
   q^2 &=& m_B^2 + m_{D^*}^2 - 2 m_B m_{D^*}\,v\cdot v' \,.
\end{eqnarray}

Equations (\ref{inelast}) and (\ref{PVff}) summarize the relations
imposed by heavy-quark symmetry on the weak decay form factors
describing the semi-leptonic decay processes $\bar B\to
D\,\ell\,\bar\nu$ and $\bar B\to D^*\ell\,\bar\nu$. These relations
are model-independent consequences of QCD in the limit where $m_b,
m_c\gg\Lambda_{\rm QCD}$. They play a crucial role in the
determination of the CKM matrix element $|V_{cb}|$. In terms of the
recoil variable $w=v\cdot v'$, the differential semi-leptonic decay
rates in the heavy-quark limit become~\cite{Vcb}
\begin{eqnarray}\label{rates}
   {{\rm d}\Gamma(\bar B\to D\,\ell\,\bar\nu)\over{\rm d}w}
   &=& {G_F^2\over 48\pi^3}\,|V_{cb}|^2\,(m_B+m_D)^2\,m_D^3\,
    (w^2-1)^{3/2}\,\xi^2(w) \,, \nonumber\\
   {{\rm d}\Gamma(\bar B\to D^*\ell\,\bar\nu)\over{\rm d}w}
   &=& {G_F^2\over 48\pi^3}\,|V_{cb}|^2\,(m_B-m_{D^*})^2\,
    m_{D^*}^3\,\sqrt{w^2-1}\,(w+1)^2 \nonumber\\
   &&\times \Bigg[ 1 + {4w\over w+1}\,
    {m_B^2 - 2 w\,m_B m_{D^*} + m_{D^*}^2\over(m_B-m_{D^*})^2}
    \Bigg]\,\xi^2(w) \,.
\end{eqnarray}
These expressions receive symmetry-breaking corrections, since the
masses of the heavy quarks are not infinitely large. Perturbative
corrections of order $\alpha_s^n(m_Q)$ can be calculated order by
order in perturbation theory. A more difficult task is to control 
the non-perturbative power corrections of order $(\Lambda_{\rm
QCD}/m_Q)^n$. The HQET provides a systematic framework for analyzing
these corrections. For the case of weak-decay form factors the
analysis of the $1/m_Q$ corrections was performed by
Luke~\cite{Luke}. Later, Falk and the present author have analyzed
the structure of $1/m_Q^2$ corrections for both meson and baryon weak
decay form factors~\cite{FaNe}. We shall not discuss these rather
technical issues in detail, but only mention the most important
result of Luke's analysis. It concerns the zero-recoil limit, where
an analogue of the Ademollo-Gatto theorem~\cite{AGTh} can be proved.
This is Luke's theorem~\cite{Luke}, which states that the matrix
elements describing the leading $1/m_Q$ corrections to weak decay
amplitudes vanish at zero recoil. This theorem is valid to all orders
in perturbation theory~\cite{FaNe,Neu7,ChGr}. Most importantly, it
protects the $\bar B\to D^*\ell\,\bar\nu$ decay rate from receiving
first-order $1/m_Q$ corrections at zero recoil~\cite{Vcb}. [A similar
statement is not true for the decay $\bar B\to D\,\ell\,\bar\nu$. The
reason is simple but somewhat subtle. Luke's theorem protects only
those form factors not multiplied by kinematic factors that vanish
for $v=v'$. By angular momentum conservation, the two pseudoscalar
mesons in the decay $\bar B\to D\,\ell\,\bar\nu$ must be in a
relative $p$ wave, and hence the amplitude is proportional to the
velocity $|\vec v_D|$ of the $D$ meson in the $B$-meson rest frame.
This leads to a factor $(w^2-1)$ in the decay rate. In such a
situation, kinematically suppressed form factors can
contribute~\cite{Neu1}.]

\subsection{Short-Distance Corrections}

In Sec.~\ref{sec:2}, we have discussed the first two steps in the
construction of the HQET. Integrating out the small components in the
heavy-quark fields, a non-local effective action was derived, which
was then expanded in a series of local operators. The effective
Lagrangian obtained that way correctly reproduces the long-distance
physics of the full theory (see Fig.~\ref{fig:magic}). It does not
contain the short-distance physics correctly, however. The reason is
obvious: a heavy quark participates in strong interactions through
its coupling to gluons. These gluons can be soft or hard, i.e.\ their
virtual momenta can be small, of the order of the confinement scale,
or large, of the order of the heavy-quark mass. But hard gluons can
resolve the spin and flavor quantum numbers of a heavy quark. Their
effects lead to a renormalization of the coefficients of the
operators in the HQET. A new feature of such short-distance
corrections is that through the running coupling constant they induce
a logarithmic dependence on the heavy-quark mass~\cite{Vol1}. Since
$\alpha_s(m_Q)$ is small, these effects can be calculated in
perturbation theory.

Consider, as an example, the matrix elements of the vector current
$V=\bar q\,\gamma^\mu Q$. In QCD this current is partially conserved
and needs no renormalization. Its matrix elements are free of 
ultraviolet divergences. Still, these matrix elements have a 
logarithmic dependence on $m_Q$ from the exchange of hard gluons 
with virtual momenta of the order of the heavy-quark mass. If one 
goes over to the effective theory by taking the limit $m_Q\to\infty$,
these logarithms diverge. Consequently, the vector current in the
effective theory does require a renormalization~\cite{PoWi}. Its
matrix elements depend on an arbitrary renormalization scale $\mu$,
which separates the regions of short- and long-distance physics. If
$\mu$ is chosen such that $\Lambda_{\rm QCD}\ll\mu\ll m_Q$, the
effective coupling constant in the region between $\mu$ and $m_Q$ is
small, and perturbation theory can be used to compute the
short-distance corrections. These corrections have to be added to the
matrix elements of the effective theory, which contain the
long-distance physics below the scale $\mu$. Schematically, then, the
relation between matrix elements in the full and in the effective
theory is
\begin{equation}\label{OPEex}
   \langle\,V(m_Q)\,\rangle_{\rm QCD}
   = C_0(m_Q,\mu)\,\langle V_0(\mu)\rangle_{\rm HQET}
   + {C_1(m_Q,\mu)\over m_Q}\,\langle V_1(\mu)\rangle_{\rm HQET}
   + \ldots \,,
\end{equation}
where we have indicated that matrix elements in the full theory
depend on $m_Q$, whereas matrix elements in the effective theory are
mass-independent, but do depend on the renormalization scale. The
Wilson coefficients $C_i(m_Q,\mu)$ are defined by this relation.
Order by order in perturbation theory, they can be computed from a
comparison of the matrix elements in the two theories. Since the
effective theory is constructed to reproduce correctly the low-energy
behavior of the full theory, this ``matching'' procedure is
independent of any long-distance physics, such as infrared
singularities, non-perturbative effects, and the nature of the 
external states used in the matrix elements.

The calculation of the coefficient functions in perturbation theory
uses the powerful methods of the renormalization group. It is in
principle straightforward, yet in practice rather tedious. A
comprehensive discussion of most of the existing calculations of
short-distance corrections in the HQET can be found in
Ref.~18.
%\citelow{review}.

\subsection{Model-Independent Determination of $|V_{cb}|$}

We will now discuss the most important application of the formalism 
described above in the context of semi-leptonic decays of $B$ mesons. 
A model-independent determination of the CKM matrix element 
$|V_{cb}|$ based on heavy-quark symmetry can be obtained by measuring
the recoil spectrum of $D^*$ mesons produced in $\bar B\to
D^*\ell\,\bar\nu$ decays~\cite{Vcb}. In the heavy-quark limit, the
differential decay rate for this process has been given in
(\ref{rates}). In order to allow for corrections to that limit, we
write
\begin{eqnarray}
   {{\rm d}\Gamma\over{\rm d}w}
   &=& {G_F^2\over 48\pi^3}\,(m_B-m_{D^*})^2\,m_{D^*}^3
    \sqrt{w^2-1}\,(w+1)^2 \nonumber\\
   &&\mbox{}\times \Bigg[ 1 + {4w\over w+1}\,
    {m_B^2-2w\,m_B m_{D^*} + m_{D^*}^2\over(m_B - m_{D^*})^2}
    \Bigg]\,|V_{cb}|^2\,{\cal{F}}^2(w) \,, 
\end{eqnarray}
where the hadronic form factor ${\cal F}(w)$ coincides with the
Isgur-Wise function up to symmetry-breaking corrections of order
$\alpha_s(m_Q)$ and $\Lambda_{\rm QCD}/m_Q$. The idea is to measure
the product $|V_{cb}|\,{\cal F}(w)$ as a function of $w$, and to
extract $|V_{cb}|$ from an extrapolation of the data to the
zero-recoil point $w=1$, where the $B$ and the $D^*$ mesons have a
common rest frame. At this kinematic point, heavy-quark symmetry
helps us to calculate the normalization ${\cal F}(1)$ with small and
controlled theoretical errors. Since the range of $w$ values
accessible in this decay is rather small ($1<w<1.5$), the
extrapolation can be done using an expansion around $w=1$:
\begin{equation}\label{Fexp}
   {\cal F}(w) = {\cal F}(1)\,\Big[ 1 - \widehat\varrho^2\,(w-1)
   + \widehat c\,(w-1)^2 \dots \Big] \,.
\end{equation}
The slope $\widehat\varrho^2$ and the curvature $\widehat c$, and 
indeed more generally the complete shape of the form factor, are
tightly constrained by analyticity and unitarity 
requirements~\cite{Boyd2,Capr}. In the long run, the statistics of 
the experimental results close to zero recoil will be such that 
these theoretical constraints will not be crucial to get a precision 
measurement of $|V_{cb}|$. They will, however, enable strong
consistency checks.

\begin{figure}[htb]
   \epsfxsize=8cm
   \vspace{0.3cm}
   \centerline{\epsffile{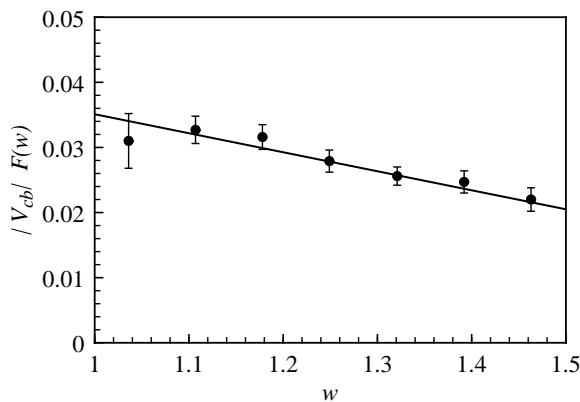}}
   \vspace{-0.3cm}
\caption{\label{fig:CLVcb}
CLEO data for the product $|V_{cb}|\,{\cal F}(w)$, as extracted from
the recoil spectrum in $\bar B\to D^*\ell\,\bar\nu$
decays~\protect\cite{CLEOVcb}. The line shows a linear fit to the
data.}
\end{figure}

Measurements of the recoil spectrum have been performed by several
experimental groups. Figure~\ref{fig:CLVcb} shows, as an example, the 
data reported some time ago by the CLEO Collaboration. The weighted 
average of the experimental results is~\cite{VcbWG}
\begin{equation}\label{VcbF}
   |V_{cb}|\,{\cal F}(1) = (35.2\pm 2.6)\times 10^{-3} \,.
\end{equation}

Heavy-quark symmetry implies that the general structure of the
symmetry-breaking corrections to the form factor at zero recoil
is~\cite{Vcb}
\begin{equation}
   {\cal F}(1) = \eta_A\,\bigg( 1 + 0\times
   {\Lambda_{\rm QCD}\over m_Q}
   + \mbox{const}\times {\Lambda_{\rm QCD}^2\over m_Q^2}
   + \dots \bigg) \equiv \eta_A\,(1+\delta_{1/m^2}) \,,
\end{equation}
where $\eta_A$ is a short-distance correction arising from the finite
renormalization of the flavor-changing axial current at zero recoil,
and $\delta_{1/m^2}$ parameterizes second-order (and higher) power
corrections. The absence of first-order power corrections at zero
recoil is a consequence of Luke's theorem~\cite{Luke}. The one-loop
expression for $\eta_A$ has been known for a long
time~\cite{Pasc,Vol2,QCD1}:
\begin{equation}\label{etaA1}
   \eta_A = 1 + {\alpha_s(M)\over\pi}\,\bigg(
   {m_b+m_c\over m_b-m_c}\,\ln{m_b\over m_c} - {8\over 3} \bigg)
   \approx 0.96 \,.
\end{equation}
The scale $M$ in the running coupling constant can be fixed by
adopting the prescription of Brodsky, Lepage and Mackenzie
(BLM)~\cite{BLM}, where it is identified with the average virtuality
of the gluon in the one-loop diagrams that contribute to $\eta_A$. If
$\alpha_s(M)$ is defined in the $\overline{\mbox{\sc ms}}$ scheme,
the result is~\cite{etaVA} $M\approx 0.51\sqrt{m_c m_b}$. Several
estimates of higher-order corrections to $\eta_A$ have been
discussed. A renormalization-group resummation of logarithms of the
type $(\alpha_s\ln m_b/m_c)^n$, $\alpha_s(\alpha_s\ln m_b/m_c)^n$ and
$m_c/m_b(\alpha_s\ln m_b/m_c)^n$ leads
to~\cite{PoWi,JiMu}$^-$\cite{QCD2} $\eta_A\approx 0.985$. On the
other hand, a resummation of ``renormalon-chain'' contributions of the 
form $\beta_0^{n-1}\alpha_s^n$, where $\beta_0=11-\frac{2}{3}n_f$ is 
the first coefficient of the QCD $\beta$-function, gives~\cite{flow}
$\eta_A\approx 0.945$. Using these partial resummations to estimate
the uncertainty gives $\eta_A = 0.965\pm 0.020$. Recently, Czarnecki
has improved this estimate by calculating $\eta_A$ at two-loop
order~\cite{Czar}. His result, $\eta_A = 0.960\pm 0.007$, is in 
excellent agreement with the BLM-improved one-loop expression
(\ref{etaA1}). Here the error is taken to be the size of the two-loop
correction.

The analysis of the power corrections is more difficult, since it
cannot rely on perturbation theory. Three approaches have been
discussed: in the ``exclusive approach'', all $1/m_Q^2$ operators in
the HQET are classified and their matrix elements estimated, leading
to~\cite{FaNe,TMann} $\delta_{1/m^2}=-(3\pm 2)\%$; the ``inclusive
approach'' has been used to derive the bound $\delta_{1/m^2}<-3\%$,
and to estimate that~\cite{Shif} $\delta_{1/m^2}=-(7\pm 3)\%$; the
``hybrid approach'' combines the virtues of the former two to obtain
a more restrictive lower bound on $\delta_{1/m^2}$. This leads
to~\cite{Vcbnew} $\delta_{1/m^2}=-0.055\pm 0.025$.

Combining the above results, adding the theoretical errors linearly
to be conservative, gives
\begin{equation}\label{F1}
   {\cal F}(1) = 0.91\pm 0.03
\end{equation}
for the normalization of the hadronic form factor at zero recoil.
Thus, the corrections to the heavy-quark limit amount to a moderate
decrease of the form factor of about 10\%. This can be used to
extract from the experimental result (\ref{VcbF}) the
model-independent value
\begin{equation}\label{Vcbexc}
   |V_{cb}| = (38.7\pm 2.8_{\rm exp}\pm 1.3_{\rm th})
   \times 10^{-3} \,.
\end{equation}

\subsection{Measurements of $\bar B\to D^*\ell\,\bar\nu$ and $\bar
B\to D\,\ell\,\bar\nu$ Form Factors and Tests of Heavy-Quark
Symmetry}

We have discussed earlier in this section that heavy-quark symmetry
implies relations between the semi-leptonic form factors of heavy
mesons. They receive symmetry-breaking corrections, which can be
estimated using the HQET. The extent to which these relations hold
can be tested experimentally by comparing the different form factors
describing the decays $\bar B\to D^{(*)}\ell\,\bar\nu$ at the same
value of $w$.

When the lepton mass is neglected, the differential decay
distributions in $\bar B\to D^*\ell\,\bar\nu$ decays can be
parameterized by three helicity amplitudes, or equivalently by three
independent combinations of form factors. It has been suggested that
a good choice for three such quantities should be inspired by the
heavy-quark limit~\cite{review,subl}. One thus defines a form factor
$h_{A1}(w)$, which up to symmetry-breaking corrections coincides with
the Isgur-Wise function, and two form-factor ratios
\begin{eqnarray}
   R_1(w) &=& \bigg[ 1 - {q^2\over(m_B+m_{D^*})^2} \bigg]\,
    {V(q^2)\over A_1(q^2)} \,, \nonumber\\
   R_2(w) &=& \bigg[ 1 - {q^2\over(m_B+m_{D^*})^2} \bigg]\,
    {A_2(q^2)\over A_1(q^2)} \,.
\end{eqnarray}
The relation between $w$ and $q^2$ has been given in (\ref{PVff}).
This definition is such that in the heavy-quark limit
$R_1(w)=R_2(w)=1$ independently of $w$.

To extract the functions $h_{A1}(w)$, $R_1(w)$ and $R_2(w)$ from
experimental data is a complicated task. However, HQET-based
calculations suggest that the $w$ dependence of the form-factor
ratios, which is induced by symmetry-breaking effects, is rather
mild~\cite{subl}. Moreover, the form factor $h_{A1}(w)$ is expected
to
have a nearly linear shape over the accessible $w$ range. This
motivates to introduce three parameters $\varrho_{A1}^2$, $R_1$ and
$R_2$ by
\begin{eqnarray}
   h_{A1}(w) &\approx& {\cal F}(1)\,\Big[ 1 - \varrho_{A1}^2 (w-1)
    \Big] \,, \nonumber\\
   R_1(w) &\approx& R_1 \,, \qquad
   R_2(w) \approx R_2 \,,
\end{eqnarray}
where ${\cal F}(1)=0.91\pm 0.03$ from (\ref{F1}). The CLEO
Collaboration has extracted these three parameters from an analysis
of the angular distributions in $\bar B\to D^*\ell\,\bar\nu$
decays~\cite{CLEff}. The results are
\begin{equation}
   \varrho_{A1}^2 = 0.91\pm 0.16 \,, \qquad
   R_1 = 1.18\pm 0.32 \,, \qquad
   R_2 = 0.71\pm 0.23 \,.
\end{equation}
Using the HQET, one obtains an essentially model-independent
prediction for the symme\-try-breaking corrections to $R_1$, whereas
the corrections to $R_2$ are somewhat model dependent. To good
approximation~\cite{review}
\begin{eqnarray}
   R_1 &\approx& 1 + {4\alpha_s(m_c)\over 3\pi}
    + {\bar\Lambda\over 2 m_c}\approx 1.3\pm 0.1 \,, \nonumber\\
   R_2 &\approx& 1 - \kappa\,{\bar\Lambda\over 2 m_c}
    \approx 0.8\pm 0.2 \,,
\end{eqnarray}
with $\kappa\approx 1$ from QCD sum rules~\cite{subl}. Here
$\bar\Lambda$ is the ``binding energy'' as defined in (\ref{Lbdef}).
Theoretical calculations~\cite{Lamsr1,Lamsr2} as well as
phenomenological analyses~\cite{GKLW,FLS2} suggest that
$\bar\Lambda\approx 0.45$--0.65~GeV is the appropriate value to be
used in one-loop calculations. A quark-model calculation of $R_1$ and
$R_2$ gives results similar to the HQET predictions~\cite{ClWa}:
$R_1\approx 1.15$ and $R_2\approx 0.91$. The experimental data
confirm the theoretical prediction that $R_1>1$ and $R_2<1$, although
the errors are still large.

Heavy-quark symmetry has also been tested by comparing the form factor 
${\cal F}(w)$ in $\bar B\to D^*\ell\,\bar\nu$ decays with the 
corresponding form factor ${\cal G}(w)$ governing $\bar B\to 
D\,\ell\,\bar\nu$ decays. The theoretical prediction~\cite{Capr,subl}
\begin{equation}
   \frac{{\cal G}(1)}{{\cal F}(1)} = 1.08\pm 0.06
\end{equation}
compares well with the experimental results for this ratio: $0.99\pm
0.19$ reported by the CLEO Collaboration~\cite{CLEOBD}, and $0.87\pm
0.30$ reported by the ALEPH Collaboration~\cite{ALEVcb}. In these
analyses, it has also been tested that within experimental errors the
shape of the two form factors agrees over the entire range of $w$
values.

The results of the analyses described above are very encouraging.
Within errors, the experiments confirm the HQET predictions, starting
to test them at the level of symmetry-breaking corrections.

\section{Inclusive Decay Rates}
\label{sec:4}

Inclusive decay rates determine the probability of the decay of a
particle into the sum of all possible final states with a given set
of global quantum numbers. An example is provided by the inclusive
semi-leptonic decay rate of the $B$ meson, $\Gamma(\bar B\to
X\,\ell\,\bar\nu)$, where the final state consists of a
lepton-neutrino pair accompanied by any number of hadrons. Here we
shall discuss the theoretical description of inclusive decays of
hadrons containing a heavy quark~\cite{Chay}$^-$\cite{MNTM}. From a
theoretical point of view such decays have two advantages: first,
bound-state effects related to the initial state, such as the ``Fermi
motion'' of the heavy quark inside the hadron~\cite{shape,Fermi}, can
be accounted for in a systematic way using the heavy-quark expansion;
secondly, the fact that the final state consists of a sum over many
hadronic channels eliminates bound-state effects related to the
properties of individual hadrons. This second feature is based on the
hypothesis of quark-hadron duality, which is an important concept in
QCD phenomenology. The assumption of duality is that cross sections
and decay rates, which are defined in the physical region (i.e.\ the
region of time-like momenta), are calculable in QCD after a
``smearing'' or ``averaging'' procedure has been applied~\cite{PQW}.
In semi-leptonic decays, it is the integration over the lepton and
neutrino phase space that provides a smearing over the invariant
hadronic mass of the final state (so-called global duality). For
non-leptonic decays, on the other hand, the total hadronic mass is
fixed, and it is only the fact that one sums over many hadronic
states that provides an averaging (so-called local duality). Clearly,
local duality is a stronger assumption than global duality. It is
important to stress that quark-hadron duality cannot yet be derived
from first principles; still, it is a necessary assumption for many
applications of QCD. The validity of global duality has been tested
experimentally using data on hadronic $\tau$ decays~\cite{Maria}.

Using the optical theorem, the inclusive decay width of a hadron
$H_b$ containing a $b$ quark can be written in the form
\begin{equation}\label{ImT}
   \Gamma(H_b\to X) = \frac{1}{m_{H_b}}\,\mbox{Im}\,
   \langle H_b|\,{\bf T}\,|H_b\rangle \,,
\end{equation}
where the transition operator ${\bf T}$ is given by
\begin{equation}
   {\bf T} = i\!\int{\rm d}^4x\,T\{\,
   {\cal L}_{\rm eff}(x),{\cal L}_{\rm eff}(0)\,\} \,.
\end{equation}
Inserting a complete set of states inside the time-ordered product,
we recover the standard expression
\begin{equation}
   \Gamma(H_b\to X) = {1\over 2 m_{H_b}}\,\sum_X\,
   (2\pi)^4\,\delta^4(p_H-p_X)\,|\langle X|\,{\cal L}_{\rm eff}\,
   |H_b\rangle|^2
\end{equation}
for the decay rate. For the case of semi-leptonic and non-leptonic
decays, ${\cal L}_{\rm eff}$ is the effective weak Lagrangian given
in (\ref{LFermi}), which in practice is corrected for short-distance
effects~\cite{AltM,Gail,cpcm3}$^-$\cite{cpcm5} arising from the
exchange of gluons with virtualities between $m_W$ and $m_b$. If some
quantum numbers of the final states $X$ are specified, the sum over
intermediate states is to be restricted appropriately. In the case of 
the inclusive semi-leptonic decay rate, for instance, the sum would
include only those states $X$ containing a lepton-neutrino pair.

\begin{figure}[htb]
   \epsfxsize=7cm
   \centerline{\epsffile{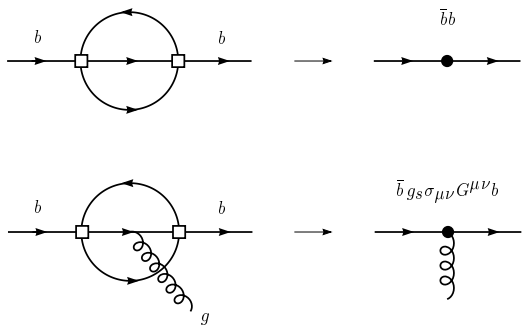}}
\caption{\label{fig:Toper}
Perturbative contributions to the transition operator ${\bf T}$
(left), and the corresponding operators in the OPE (right). The open
squares represent a four-fermion interaction of the effective
Lagrangian ${\cal L}_{\rm eff}$, while the black circles represent
local operators in the OPE.}
\end{figure}

In perturbation theory, some contributions to the transition operator
are given by the two-loop diagrams shown on the left-hand side in
Fig.~\ref{fig:Toper}. Because of the large mass of the $b$ quark, the
momenta flowing through the internal propagator lines are large. It
is thus possible to construct an OPE for the transition operator, in
which ${\bf T}$ is represented as a series of local operators
containing the heavy-quark fields. The operator with the lowest
dimension, $d=3$, is $\bar b b$. It arises by contracting the
internal lines of the first diagram. The only gauge-invariant
operator with dimension 4 is $\bar b\,i\rlap{\,/}D\,b$; however, the
equations of motion imply that between physical states this operator
can be replaced by $m_b\bar b b$. The first operator that is
different from $\bar b b$ has dimension 5 and contains the gluon
field. It is given by $\bar b\,g_s\sigma_{\mu\nu} G^{\mu\nu} b$. This
operator arises from diagrams in which a gluon is emitted from one of
the internal lines, such as the second diagram shown in
Fig.~\ref{fig:Toper}. For dimensional reasons, the matrix elements of
such higher-dimensional operators are suppressed by inverse powers of
the heavy-quark mass. Thus, any inclusive decay rate of a hadron
$H_b$ can be written as~\cite{Bigi}$^-$\cite{MaWe}
\begin{equation}\label{gener}
   \Gamma(H_b\to X_f) = {G_F^2 m_b^5\over 192\pi^3}\,
   \bigg\{ c_3^f\,\langle\bar b b\rangle_H
   + c_5^f\,{\langle\bar b\,g_s\sigma_{\mu\nu} G^{\mu\nu} b
   \rangle_H\over m_b^2} + \dots \bigg\} \,,
\end{equation}
where the prefactor arises naturally from the loop integrations,
$c_n^f$ are calculable coefficient functions (which also contain the
relevant CKM matrix elements) depending on the quantum numbers $f$ of
the final state, and $\langle O\rangle_H$ are the (normalized)
forward matrix elements of local operators, for which we use the
short-hand notation
\begin{equation}
   \langle O\rangle_H = {1\over 2 m_{H_b}}\,\langle H_b|\,
   O\,|H_b\rangle \,.
\end{equation}

In the next step, these matrix elements are systematically expanded
in powers of $1/m_b$, using the technology of the HQET. The result
is~\cite{FaNe,Bigi,MaWe}
\begin{eqnarray}
   \langle\bar b b\rangle_H &=& 1
    - {\mu_\pi^2(H_b)-\mu_G^2(H_b)\over 2 m_b^2} + O(1/m_b^3) \,,
    \nonumber\\
   \langle\bar b\,g_s\sigma_{\mu\nu} G^{\mu\nu} b\rangle_H
   &=& 2\mu_G^2(H_b) + O(1/m_b) \,,
\end{eqnarray}
where we have defined the HQET matrix elements
\begin{eqnarray}
   \mu_\pi^2(H_b) &=& {1\over 2 m_{H_b}}\,
    \langle H_b(v)|\,\bar b_v\,(i\vec D)^2\,b_v\,|H_b(v)\rangle \,,
    \nonumber\\
   \mu_G^2(H_b) &=& {1\over 2 m_{H_b}}\,
    \langle H_b(v)|\,\bar b_v {g_s\over 2}\sigma_{\mu\nu}
     G^{\mu\nu} b_v\,|H_b(v)\rangle \,.
\end{eqnarray}
Here $(i\vec D)^2=(i v\cdot D)^2-(i D)^2$; in the rest frame, this is
the square of the operator for the spatial momentum of the heavy
quark. Inserting these results into (\ref{gener}) yields
\begin{equation}\label{generic}
   \Gamma(H_b\to X_f) = {G_F^2 m_b^5\over 192\pi^3}\,
   \bigg\{ c_3^f\,\bigg( 1
   - {\mu_\pi^2(H_b)-\mu_G^2(H_b)\over 2 m_b^2} \bigg)
   + 2 c_5^f\,{\mu_G^2(H_b)\over m_b^2} + \dots \bigg\} \,.
\end{equation}
It is instructive to understand the appearance of the ``kinetic
energy'' contribution $\mu_\pi^2$, which is the gauge-covariant
extension of the square of the $b$-quark momentum inside the heavy
hadron. This contribution is the field-theory analogue of the Lorentz
factor $(1-\vec v_b^{\,2})^{1/2}\simeq 1-\vec k^{\,2}/2 m_b^2$, in
accordance with the fact that the lifetime, $\tau=1/\Gamma$, for a
moving particle increases due to time dilation.

The main result of the heavy-quark expansion for inclusive decay
rates is the observation that the free quark decay (i.e.\ the parton
model) provides the first term in a systematic $1/m_b$ expansion
\cite{Chay}. For dimensional reasons, the corresponding rate is
proportional to the fifth power of the $b$-quark mass. The
non-perturbative corrections, which arise from bound-state effects
inside the $B$ meson, are suppressed by at least two powers of the
heavy-quark mass, i.e.\ they are of relative order $(\Lambda_{\rm
QCD}/m_b)^2$. Note that the absence of first-order power corrections
is a consequence of the equations of motion, as there is no
independent gauge-invariant operator of dimension 4 that could appear
in the OPE. The fact that bound-state effects in inclusive decays are
strongly suppressed explains a posteriori the success of the parton
model in describing such processes~\cite{ACCMM,Pasch}.

The hadronic matrix elements appearing in the heavy-quark expansion
(\ref{generic}) can be determined to some extent from the known
masses of heavy hadron states. For the $B$ meson, one finds that
\begin{eqnarray}\label{mupimuG}
   \mu_\pi^2(B) &=& - \lambda_1 = (0.3\pm 0.2)~\mbox{GeV}^2 \,,
    \nonumber\\
   \mu_G^2(B) &=& 3\lambda_2\approx 0.36~\mbox{GeV}^2 \,,
\end{eqnarray}
where $\lambda_1$ and $\lambda_2$ are the parameters appearing in the
mass formula (\ref{FNrela}). For the ground-state baryon $\Lambda_b$,
in which the light constituents have total spin zero, it follows that
\begin{equation}
   \mu_G^2(\Lambda_b) = 0 \,,
\end{equation}
while the matrix element $\mu_\pi^2(\Lambda_b)$ obeys the relation
\begin{equation}
   (m_{\Lambda_b}-m_{\Lambda_c}) - (\overline{m}_B-\overline{m}_D)
   = \Big[ \mu_\pi^2(B)-\mu_\pi^2(\Lambda_b) \Big]\,\bigg(
   {1\over 2 m_c} - {1\over 2 m_b} \bigg) + O(1/m_Q^2) \,,
\end{equation}
where $\overline{m}_B$ and $\overline{m}_D$ denote the spin-averaged
masses introduced in connection with (\ref{mbmc}). The above relation 
implies
\begin{equation}\label{mupidif}
   \mu_\pi^2(B) - \mu_\pi^2(\Lambda_b) = (0.01\pm 0.03)~\mbox{GeV}^2
   \,.
\end{equation}
What remains to be calculated, then, is the coefficient functions
$c_n^f$ for a given inclusive decay channel. 

To illustrate this general formalism, we discuss as an example the 
determination of $|V_{cb}|$ from inclusive semi-leptonic $B$ decays.
In this case the short-distance coefficients in the general 
expression (\ref{generic}) are given by~\cite{Bigi}$^-$\cite{MaWe}
\begin{eqnarray}
   c_3^{\rm SL} &=& |V_{cb}|^2 \Big[ 1 - 8 x^2 + 8 x^6 - x^8
    - 12 x^4\ln x^2 + O(\alpha_s) \Big] \,, \nonumber\\
   c_5^{\rm SL} &=& -6 |V_{cb}|^2 (1-x^2)^4 \,.
\end{eqnarray}
Here $x=m_c/m_b$, and $m_b$ and $m_c$ are the masses of the $b$
and $c$ quarks, defined to a given order in perturbation theory
\cite{Tarr}. The $O(\alpha_s)$ terms in $c_3^{\rm SL}$ are known
exactly~\cite{fgrefs}, and reliable estimates exist for the 
$O(\alpha_s^2)$ corrections \cite{Oal2}. The theoretical
uncertainties in this determination of $|V_{cb}|$ are quite different
from those entering the analysis of exclusive decays. The main
sources are the dependence on the heavy-quark masses, 
higher-order perturbative corrections, and above all the assumption 
of global quark-hadron duality. A conservative estimate of the total
theoretical error on the extracted value of $|V_{cb}|$ 
yields~\cite{BaBar}
\begin{equation}
   |V_{cb}| = (0.040\pm 0.003)
   \bigg[ {\hbox{B}_{\rm SL}\over 10.5\%} \bigg]^{1/2}
   \bigg[ \frac{1.6\,\mbox{ps}}{\tau_B} \bigg]^{1/2}
   = (40\pm 1_{\rm exp}\pm 3_{\rm th}) \times 10^{-3} \,.
\end{equation}
The value of $|V_{cb}|$ extracted from the inclusive semi-leptonic 
width is in excellent agreement with the value in (\ref{Vcbexc}) 
obtained from the analysis of the exclusive decay 
$\bar B\to D^*\ell\,\bar\nu$. This agreement is gratifying given the 
differences of the methods used, and it provides an indirect test of 
global quark-hadron duality. Combining the two measurements gives the 
final result
\begin{equation}
   |V_{cb}| = 0.039\pm 0.002 \,.
\end{equation}
After $V_{ud}$ and $V_{us}$, this is the third-best known entry in
the CKM matrix.

\section{\boldmath Rare $B$ Decays and Determination of the Weak 
Phase $\gamma$\unboldmath}
\label{sec:5}

The main objectives of the $B$ factories are to explore the physics 
of CP violation, to determine the flavor parameters of the 
electroweak theory, and to probe for physics beyond the 
Standard Model. This will test the CKM mechanism, which predicts that 
all CP violation results from a single complex phase in the quark 
mixing matrix. Facing the announcement of evidence for a CP asymmetry 
in the decays $B\to J/\psi\,K_S$ by the CDF Collaboration~\cite{CDF}, 
the confirmation of direct CP violation in $K\to\pi\pi$ decays by the 
KTeV and NA48 groups~\cite{KTeV,NA48}, and the successful start of 
the $B$ factories at SLAC and KEK, the year 1999 has been an important 
step towards achieving this goal.

\begin{figure}[htb]
   \epsfxsize=6cm
   \centerline{\epsffile{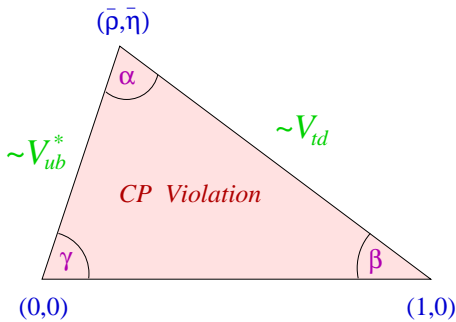}}
\caption{\label{fig:UT}
The rescaled unitarity triangle representing the relation 
$1+\frac{V_{ub}^* V_{ud}}{V_{cb}^* V_{cd}}
+\frac{V_{tb}^* V_{td}}{V_{cb}^* V_{cd}}=0$. The apex is determined 
by the Wolfenstein parameters $(\bar\rho,\bar\eta)$. The area of the
triangle is proportional to the strength of CP violation in the 
Standard Model.}
\end{figure}

The determination of the sides and angles of the ``unitarity 
triangle'' $V_{ub}^* V_{ud}+V_{cb}^* V_{cd}+V_{tb}^* V_{td}=0$ 
depicted in Fig.~\ref{fig:UT} plays a central role in the $B$ factory 
program. Adopting the standard phase conventions for the CKM matrix, 
only the two smallest elements in this relation, $V_{ub}^*$ and 
$V_{td}$, have non-vanishing imaginary parts (to an excellent 
approximation). In the Standard Model the angle 
$\beta=-\mbox{arg}(V_{td})$ can be determined in a theoretically clean 
way by measuring the mixing-induced CP asymmetry in the decays 
$B\to J/\psi\,K_S$. The preliminary CDF result implies~\cite{CDF} 
$\sin2\beta=0.79_{-0.44}^{+0.41}$. The angle 
$\gamma=\mbox{arg}(V_{ub}^*)$, or equivalently the combination 
$\alpha=180^\circ-\beta-\gamma$, is much harder to 
determine~\cite{BaBar}. Recently, there has been significant progress 
in the theoretical understanding of the hadronic decays $B\to\pi K$, 
and methods have been developed to extract information on $\gamma$ 
from rate measurements for these processes. Here we discuss the 
charged modes $B^\pm\to\pi K$, which from a theoretical perspective 
are particularly clean.

In the Standard Model, the main contributions to the decay amplitudes 
for the rare processes $B\to\pi K$ are due to the penguin-induced 
flavor-changing neutral current (FCNC) transitions 
$\bar b\to\bar s q\bar q$, which exceed a small, Cabibbo-suppressed 
$\bar b\to\bar u u\bar s$ contribution from $W$-boson exchange. The 
weak phase $\gamma$ enters through the interference of these two 
(``penguin'' and ``tree'') contributions. Because of a fortunate 
interplay of isospin, Fierz and flavor symmetries, the theoretical 
description of the charged modes $B^\pm\to\pi K$ is very clean 
despite the fact that these are exclusive non-leptonic 
decays~\cite{us}$^-$\cite{me}. Without any dynamical assumption, 
the hadronic uncertainties in the description of the interference 
terms relevant to the determination of $\gamma$ are of relative 
magnitude $O(\lambda^2)$ or $O(\epsilon_{\rm SU(3)}/N_c)$, where 
$\lambda=\sin\theta_C\approx 0.22$ is a measure of Cabibbo 
suppression, $\epsilon_{\rm SU(3)}\sim 20\%$ is the typical size 
of SU(3) breaking, and the factor $1/N_c$ indicates that the 
corresponding terms vanish in the factorization approximation. 
Factorizable SU(3) breaking can be accounted for in a straightforward 
way. 

Recently, the accuracy of this description has been further improved 
when it was shown that non-leptonic $B$ decays into two light mesons, 
such as $B\to\pi K$ and $B\to\pi\pi$, admit a systematic heavy-quark 
expansion~\cite{fact}. To leading order in $1/m_b$, but to all 
orders in perturbation theory, the decay amplitudes for these 
processes can be calculated from first principles without recourse 
to phenomenological models. The QCD factorization theorem proved 
in Ref.~122
%\citelow{fact} 
improves upon the phenomenological approach of 
``generalized factorization''~\cite{BStech}, which emerges as the 
leading term in the heavy-quark limit. With the help of this theorem, 
the irreducible theoretical uncertainties in the description of the 
$B^\pm\to\pi K$ decay amplitudes can be reduced by an extra factor of 
$O(1/m_b)$, rendering their analysis essentially model 
independent. As a consequence of this fact, and because they are 
dominated by FCNC transitions, the decays $B^\pm\to\pi K$ offer a 
sensitive probe to physics beyond the Standard 
Model~\cite{me,Mati}$^-$\cite{troja}, much in the same way as the 
``classical'' FCNC processes $B\to X_s\gamma$ or 
$B\to X_s\,\ell^+\ell^-$. 

\subsection{Theory of $B^\pm\to\pi K$ Decays}

The hadronic decays $B\to\pi K$ are mediated by a low-energy
effective weak Hamiltonian~\cite{Heff}, whose operators allow for 
three different classes of flavor topologies: QCD penguins, trees, 
and electroweak penguins. In the Standard Model the weak couplings 
associated with these topologies are known. From the measured 
branching ratios one can deduce that QCD penguins dominate the 
$B\to\pi K$ decay amplitudes~\cite{Digh}, whereas trees and 
electroweak penguins are subleading and of a similar 
strength~\cite{oldDesh}. The theoretical description of the two 
charged modes $B^\pm\to\pi^\pm K^0$ and $B^\pm\to\pi^0 K^\pm$ 
exploits the fact that the amplitudes for these processes differ in 
a pure isospin amplitude, $A_{3/2}$, defined as the matrix element of 
the isovector part of the effective Hamiltonian between a $B$ meson 
and the $\pi K$ isospin eigenstate with $I=\frac 32$. In the Standard 
Model the parameters of this amplitude are determined, up to an 
overall strong phase $\phi$, in the limit of SU(3) flavor 
symmetry~\cite{us}. Using the QCD factorization theorem the 
SU(3)-breaking corrections can be calculated in a model-independent 
way up to non-factorizable terms that are power-suppressed in $1/m_b$ 
and vanish in the heavy-quark limit. 

A convenient parameterization of the non-leptonic decay amplitudes 
${\cal A}_{+0}\equiv{\cal A}(B^+\to\pi^+ K^0)$ and
${\cal A}_{0+}\equiv -\sqrt2\,{\cal A}(B^+\to\pi^0 K^+)$ is~\cite{me}
\begin{eqnarray}\label{ampls}
   {\cal A}_{+0} &=& P\,(1-\varepsilon_a\,e^{i\gamma} e^{i\eta})
    \,, \nonumber\\
   {\cal A}_{0+} &=& P \Big[ 1 - \varepsilon_a\,e^{i\gamma}
    e^{i\eta} - \varepsilon_{3/2}\,e^{i\phi}
    (e^{i\gamma} - \delta_{\rm EW}) \Big] \,,
\end{eqnarray}
where $P$ is the dominant penguin amplitude defined as the sum 
of all terms in the $B^+\to\pi^+ K^0$ amplitude not proportional 
to $e^{i\gamma}$, $\eta$ and $\phi$ are strong phases, and 
$\varepsilon_a$, $\varepsilon_{3/2}$ and $\delta_{\rm EW}$ are 
real hadronic parameters. The weak phase $\gamma$ changes sign 
under a CP transformation, whereas all other parameters stay 
invariant. 

Based on a naive quark-diagram analysis one would not expect the 
$B^+\to\pi^+ K^0$ amplitude to receive a contribution from 
$\bar b\to\bar u u\bar s$ tree topologies; however, such a 
contribution can be induced through final-state rescattering or 
annihilation contributions~\cite{BBlok}$^-$\cite{At97}. They are 
parameterized by $\varepsilon_a=O(\lambda^2)$. In the heavy-quark 
limit this parameter can be calculated and is found to be very 
small~ \cite{newfact}: $\varepsilon_a\approx-2\%$. In the future, 
it will be possible to put upper and lower bounds on 
$\varepsilon_a$ by comparing the CP-averaged branching ratios for 
the decays~\cite{Fa97} $B^\pm\to\pi^\pm K^0$ and 
$B^\pm\to K^\pm\bar K^0$. Below we assume $|\varepsilon_a|\le 0.1$; 
however, our results will be almost insensitive to this assumption.

The terms proportional to $\varepsilon_{3/2}$ in (\ref{ampls}) 
parameterize the isospin amplitude $A_{3/2}$. The weak phase 
$e^{i\gamma}$ enters through the tree process 
$\bar b\to\bar u u\bar s$, whereas the quantity $\delta_{\rm EW}$ 
describes the effects of electroweak penguins. The parameter 
$\varepsilon_{3/2}$ measures the relative strength of tree and 
QCD penguin contributions. Information about it can be derived by 
using SU(3) flavor symmetry to relate the tree contribution to 
the isospin amplitude $A_{3/2}$ to the corresponding contribution 
in the decay $B^+\to\pi^+\pi^0$. Since the final state $\pi^+\pi^0$ 
has isospin $I=2$, the amplitude for this process does not receive 
any contribution from QCD penguins. Moreover, electroweak penguins 
in $\bar b\to\bar d q\bar q$ transitions are negligibly small. We 
define a related parameter $\bar\varepsilon_{3/2}$ by writing 
\begin{equation}
   \varepsilon_{3/2} = \bar\varepsilon_{3/2}
   \sqrt{1-2\varepsilon_a\cos\eta\cos\gamma+\varepsilon_a^2} \,,
\end{equation}
so that the two quantities agree in the limit $\varepsilon_a\to 0$. 
In the SU(3) limit this new parameter can be determined 
experimentally form the relation~\cite{us}
\begin{equation}\label{eps}
   \bar\varepsilon_{3/2} = R_1
   \left|\frac{V_{us}}{V_{ud}}\right| \left[
   \frac{2\mbox{B}(B^\pm\to\pi^\pm\pi^0)}
        {\mbox{B}(B^\pm\to\pi^\pm K^0)} \right]^{1/2} .
\end{equation}
SU(3)-breaking corrections are described by the factor 
$R_1=1.22\pm 0.05$, which can be calculated in a model-independent 
way using the QCD factorization theorem for non-leptonic 
decays~\cite{newfact}. The quoted error is an estimate of the 
theoretical uncertainty due to corrections of 
$O(\frac{1}{N_c}\frac{m_s}{m_b})$. Using preliminary data reported by 
the CLEO Collaboration~\cite{CLEO} to evaluate the ratio of the 
CP-averaged branching ratios in (\ref{eps}), we obtain
\begin{equation}\label{epsval}
   \bar\varepsilon_{3/2} = 0.21\pm 0.06_{\rm exp}
   \pm 0.01_{\rm th} \,.
\end{equation}
With a better measurement of the branching ratios the uncertainty 
in $\bar\varepsilon_{3/2}$ will be reduced significantly.

Finally, the parameter
\begin{eqnarray}\label{delta}
   \delta_{\rm EW} &=& R_2\,
    \left| \frac{V_{cb}^* V_{cs}}{V_{ub}^* V_{us}} \right|\,
    \frac{\alpha}{8\pi}\,\frac{x_t}{\sin^2\!\theta_W}
    \left( 1 + \frac{3\ln x_t}{x_t-1} \right) \nonumber\\
   &=& (0.64\pm 0.09)\times\frac{0.085}{|V_{ub}/V_{cb}|} \,,
\end{eqnarray}
with $x_t=(m_t/m_W)^2$, describes the ratio of electroweak penguin 
and tree contributions to the isospin amplitude $A_{3/2}$. In the 
SU(3) limit it is calculable in terms of Standard Model 
parameters~\cite{us,Fl96}. SU(3)-breaking as well as small 
electromagnetic corrections are accounted for by the 
quantity~\cite{me,newfact} $R_2=0.92\pm 0.09$. The error quoted in 
(\ref{delta}) includes the uncertainty in the top-quark mass. 

Important observables in the study of the weak phase $\gamma$ are 
the ratio of the CP-averaged branching ratios in the two 
$B^\pm\to\pi K$ decay modes,
\begin{equation}\label{Rst}
   R_* = \frac{\mbox{B}(B^\pm\to\pi^\pm K^0)}
              {2\mbox{B}(B^\pm\to\pi^0 K^\pm)} 
   = 0.75\pm 0.28 \,,
\end{equation}
and a particular combination of the direct CP asymmetries,
\begin{equation}\label{Atil}
   \widetilde A = \frac{A_{\rm CP}(B^\pm\to\pi^0 K^\pm)}{R_*}
    - A_{\rm CP}(B^\pm\to\pi^\pm K^0)
   = -0.52\pm 0.42 \,.
\end{equation}
The experimental values of these quantities are derived using
preliminary data reported by the CLEO Collaboration~\cite{CLEO}. 
The theoretical expressions for $R_*$ and $\widetilde A$ obtained 
using the parameterization in (\ref{ampls}) are
\begin{eqnarray}\label{expr}
   R_*^{-1} &=& 1 + 2\bar\varepsilon_{3/2}\cos\phi\,
    (\delta_{\rm EW}-\cos\gamma) \nonumber\\
   &&\mbox{}+ \bar\varepsilon_{3/2}^2
    (1-2\delta_{\rm EW}\cos\gamma+\delta_{\rm EW}^2)
    \!+\! O(\bar\varepsilon_{3/2}\varepsilon_a) \,, \nonumber\\
   \widetilde A &=& 2\bar\varepsilon_{3/2} \sin\gamma \sin\phi
    + O(\bar\varepsilon_{3/2}\varepsilon_a) \,.
\end{eqnarray}
Note that the rescattering effects described by $\varepsilon_a$ 
are suppressed by a factor of $\bar\varepsilon_{3/2}$ and thus 
reduced to the percent level. Explicit expressions for these
contributions can be found in Ref.~121.
%\citelow{me}.

\subsection{Lower Bound on $\gamma$ and Constraint in the 
$(\bar\rho,\bar\eta)$ Plane}

There are several strategies for exploiting the above relations. 
From a measurement of the ratio $R_*$ alone a bound on $\cos\gamma$ 
can be derived, implying a non-trivial constraint on the Wolfenstein 
parameters $\bar\rho$ and $\bar\eta$ defining the apex of the 
unitarity triangle~\cite{us}. Only CP-averaged branching ratios are 
needed for this purpose. Varying the strong phases $\phi$ and $\eta$ 
independently we first obtain an upper bound on the inverse of 
$R_*$. Keeping terms of linear order in $\varepsilon_a$ 
yields~\cite{me}
\begin{equation}\label{Rbound}
   R_*^{-1} \le \left( 1 + \bar\varepsilon_{3/2}\,
   |\delta_{\rm EW}-\cos\gamma| \right)^2
   + \bar\varepsilon_{3/2}^2\sin^2\!\gamma
   + 2\bar\varepsilon_{3/2}|\varepsilon_a|\sin^2\!\gamma \,.
\end{equation}
Provided $R_*$ is significantly smaller than 1, this bound implies 
an exclusion region for $\cos\gamma$ which becomes larger the 
smaller the values of $R_*$ and $\bar\varepsilon_{3/2}$ are. It is 
convenient to consider instead of $R_*$ the related 
quantity~\cite{troja}
\begin{equation}\label{Rval}
   X_R = \frac{\sqrt{R_*^{-1}}-1}{\bar\varepsilon_{3/2}} 
   = 0.72\pm 0.98_{\rm exp}\pm 0.03_{\rm th} \,.
\end{equation}
Because of the theoretical factor $R_1$ entering the definition of 
$\bar\varepsilon_{3/2}$ in (\ref{eps}) this is, strictly speaking, 
not an observable. However, the theoretical uncertainty in $X_R$ is 
so much smaller than the present experimental error that it can be
ignored for all practical purposes. The advantage of presenting our 
results in terms of $X_R$ rather than $R_*$ is that the leading 
dependence on $\bar\varepsilon_{3/2}$ cancels out, leading to the 
simple bound $|X_R|\le|\delta_{\rm EW}-\cos\gamma|
+O(\bar\varepsilon_{3/2},\varepsilon_a)$.

\begin{figure}[htb]
   \epsfxsize=7cm
   \centerline{\epsffile{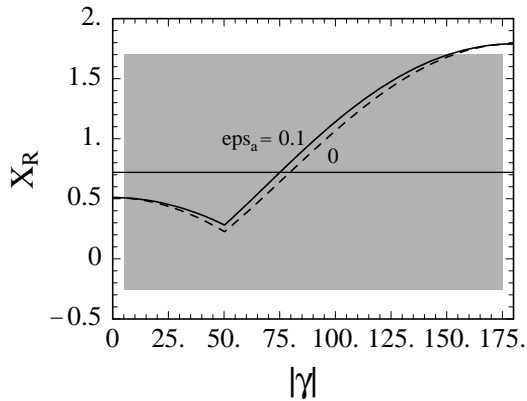}}
\caption{\label{fig:Rbound}
Theoretical upper bound on the ratio $X_R$ versus $|\gamma|$ for 
$\varepsilon_a=0.1$ (solid line) and $\varepsilon_a=0$ (dashed 
line). The horizontal line and band show the current experimental 
value with its $1\sigma$ variation.}
\end{figure}

In Fig.~\ref{fig:Rbound} we show the upper bound on $X_R$ as a 
function of $|\gamma|$, obtained by varying the input parameters in 
the intervals $0.15\le\bar\varepsilon_{3/2}\le 0.27$ and 
$0.49\le\delta_{\rm EW}\le 0.79$ (corresponding to using 
$|V_{ub}/V_{cb}|=0.085\pm 0.015$ in (\ref{delta})). Note that the 
effect of the rescattering contribution parameterized by 
$\varepsilon_a$ is very small. The gray band shows the current 
value of $X_R$, which still has too large an error to provide any 
useful information on $\gamma$. The situation may change, however, 
once a more precise measurement of $X_R$ will become available. For 
instance, if the current central value $X_R=0.72$ were confirmed, it 
would imply the bound $|\gamma|>75^\circ$, marking a significant 
improvement over the indirect limit $|\gamma|>37^\circ$ inferred from 
the global analysis of the unitarity triangle including information 
from $K$--$\bar K$ mixing~\cite{BaBar}. 

\begin{figure}[htb]
   \epsfxsize=9cm
   \centerline{\epsffile{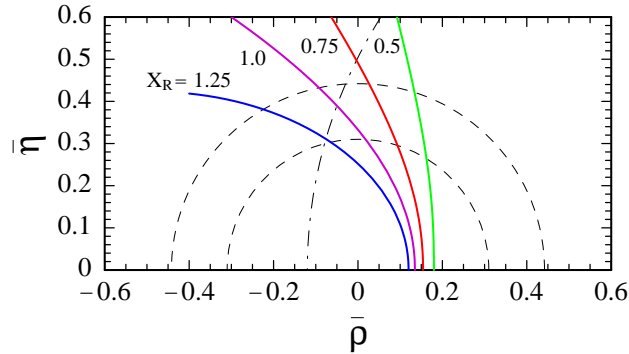}}
\caption{\label{fig:CKMbound}
Theoretical constraints on the Wolfenstein parameters 
$(\bar\rho,\bar\eta)$ implied by a measurement of the ratio $X_R$ in 
$B^\pm\to\pi K$ decays (solid lines), semi-leptonic $B$ decays (dashed 
circles), and $B_{d,s}$--$\bar B_{d,s}$ mixing (dashed-dotted line).}
\end{figure}

So far, we have used the inequality (\ref{Rbound}) to derive a lower 
bound on $|\gamma|$. However, a large part of the uncertainty in the 
value of $\delta_{\rm EW}$, and thus in the resulting bound on 
$|\gamma|$, comes from the present large error on $|V_{ub}|$. Since 
this is not a hadronic uncertainty, it is appropriate to separate it 
and turn (\ref{Rbound}) into a constraint on the Wolfenstein 
parameters $\bar\rho$ and $\bar\eta$. To this end, we use that 
$\cos\gamma=\bar\rho/\sqrt{\bar\rho^2+\bar\eta^2}$ by definition, 
and $\delta_{\rm EW}=(0.24\pm 0.03)/\sqrt{\bar\rho^2+\bar\eta^2}$ 
from (\ref{delta}). The solid lines in Fig.~\ref{fig:CKMbound} 
show the resulting constraint in the $(\bar\rho,\bar\eta)$ plane 
obtained for the representative values $X_R=0.5$, 0.75, 1.0, 1.25 
(from right to left), which for $\bar\varepsilon_{3/2}=0.21$
would correspond to $R_*=0.82$, 0.75, 0.68, 0.63, respectively. 
Values to the right of these lines are excluded. For comparison, the 
dashed circles show the constraint arising from the measurement of 
the ratio $|V_{ub}/V_{cb}|=0.085\pm 0.015$ in semi-leptonic $B$ 
decays, and the dashed-dotted line shows the bound implied by the 
present experimental limit on the mass difference $\Delta m_s$ in the 
$B_s$ system \cite{BaBar}. Values to the left of this line are 
excluded. It is evident from the figure that the bound resulting from 
a measurement of the ratio $X_R$ in $B^\pm\to\pi K$ decays may be 
very non-trivial and, in particular, may eliminate the possibility 
that $\gamma=0$. The combination of this bound with information from 
semi-leptonic decays and $B$--$\bar B$ mixing alone would then 
determine the Wolfenstein parameters $\bar\rho$ and $\bar\eta$ 
within narrow ranges,\footnote{An observation of CP violation, such 
as the measurement of $\epsilon_K$ in $K$--$\bar K$ mixing or 
$\sin2\beta$ in $B\to J/\psi\,K_S$ decays, is however needed to 
fix the sign of $\bar\eta$.}
and in the context of the CKM model would prove the existence of 
direct CP violation in $B$ decays. If one is more optimistic, one may 
even hope that in the future the constraint from $B\to\pi K$ decays 
may become incompatible with the bound from $B_s$--$\bar B_s$ mixing,
thus indicating New Physics beyond the Standard Model.\footnote{At 
the time of writing, the bound from $B_s$--$\bar B_s$ mixing is being 
pushed further to the right, making such a scenario a tantalizing 
possibility.}

\subsection{Extraction of $\gamma$}

Ultimately, the goal is of course not only to derive a bound on 
$\gamma$ but to determine this parameter directly from the data. 
This requires to fix the strong phase $\phi$ in (\ref{expr}), which 
can be achieved either through the measurement of a CP asymmetry or 
with the help of theory. A strategy for an experimental determination 
of $\gamma$ from $B^\pm\to\pi K$ decays has been suggested 
in Ref.~120.
%\citelow{us2} 
It generalizes a method proposed by Gronau, Rosner and 
London~\cite{GRL} to include the effects of electroweak penguins. The 
approach has later been refined to account for rescattering 
contributions to the $B^\pm\to\pi^\pm K^0$ decay amplitudes~\cite{me}. 
Before discussing this method, we will first illustrate an 
easier strategy for a theory-guided determination of $\gamma$ based 
on the QCD factorization theorem for non-leptonic decays~\cite{fact}. 
This method does not require any measurement of a CP asymmetry.

\vspace{0.2cm}\noindent
{\em Theory-guided determination:}\\
In the previous section the theoretical predictions for the 
non-leptonic $B\to\pi K$ decay amplitudes obtained using the QCD 
factorization theorem were used in a minimal way, i.e.\ only to 
calculate the size of the SU(3)-breaking effects parameterized by 
$R_1$ and $R_2$ in (\ref{eps}) and (\ref{delta}). The resulting bound 
on $\gamma$ and the corresponding constraint in the 
$(\bar\rho,\bar\eta)$ plane are therefore theoretically very clean. 
However, they are only useful if the value of $X_R$ is found to be 
larger than about 0.5 (see Fig.~\ref{fig:Rbound}), in which case 
values of $|\gamma|$ below $65^\circ$ are excluded. If it would turn 
out that $X_R<0.5$, then it is possible to satisfy the inequality 
(\ref{Rbound}) also for small values of $\gamma$, however, at the 
price of having a very large strong phase, $\phi\approx 180^\circ$. 
But this possibility can be discarded based on the model-independent 
prediction that~\cite{fact}
\begin{equation}\label{phiest}
   \phi = O[\alpha_s(m_b),\Lambda_{\rm QCD}/m_b] \,.
\end{equation}
A direct calculation of this phase to leading power in $1/m_b$ 
yields~\cite{newfact} $\phi\approx-11^\circ$. Using the fact that 
$\phi$ is parametrically small, we can exploit a measurement of the 
ratio $X_R$ to obtain a determination of $|\gamma|$ -- corresponding 
to an allowed region in the $(\bar\rho,\bar\eta)$ plane -- rather 
than just a bound. This determination is unique up to a sign. Note that 
for small values of $\phi$ the impact of the strong phase in the 
expression for $R_*$ in (\ref{expr}) is a second-order effect. As 
long as $|\phi|\ll\sqrt{2\Delta\bar\varepsilon_{3/2}/
\bar\varepsilon_{3/2}}$, the uncertainty in $\cos\phi$ has a much 
smaller effect than the uncertainty in $\bar\varepsilon_{3/2}$. With 
the present value of $\bar\varepsilon_{3/2}$ this is the case as long 
as $|\phi|\ll 43^\circ$. We believe it is a safe assumption to take 
$|\phi|<25^\circ$ (i.e.\ more than twice the value obtained to 
leading order in $1/m_b$), so that $\cos\phi>0.9$. 

\begin{figure}[htb]
   \epsfxsize=10.5cm
   \centerline{\epsffile{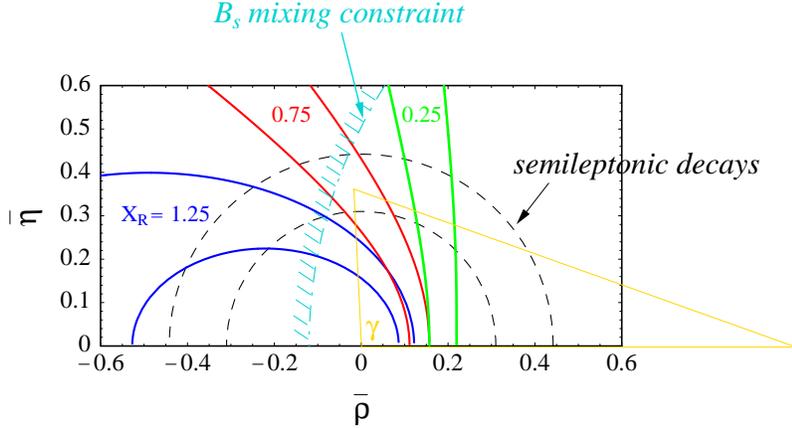}}
\caption{\label{fig:CKMfit}
Allowed regions in the $(\bar\rho,\bar\eta)$ plane for fixed values 
of $X_R$, obtained by varying all theoretical parameters inside 
their respective ranges of uncertainty, as specified in the text. 
The sign of $\bar\eta$ is not determined.}
\end{figure}

Solving the equation for $R_*$ in (\ref{expr}) for $\cos\gamma$, and 
including the corrections of $O(\varepsilon_a)$, we find 
\begin{equation}\label{gamth}
   \cos\gamma = \delta_{\rm EW}
    - \frac{X_R + \frac12\bar\varepsilon_{3/2}
            (X_R^2-1+\delta_{\rm EW}^2)}
           {\cos\phi+\bar\varepsilon_{3/2}\delta_{\rm EW}}
    + \frac{\varepsilon_a\cos\eta\sin^2\!\gamma}
           {\cos\phi+\bar\varepsilon_{3/2}\delta_{\rm EW}} \,,
\end{equation}
where we have set $\cos\phi=1$ in the numerator of the 
$O(\varepsilon_a)$ term. Using the QCD factorization theorem one 
finds that $\varepsilon_a\cos\eta\approx -0.02$ in the heavy-quark 
limit~\cite{newfact}, and we assign a 100\% uncertainty to this 
estimate. In evaluating the result (\ref{gamth}) we scan the 
parameters in the ranges $0.15\le\bar\varepsilon_{3/2}\le 0.27$, 
$0.55\le\delta_{\rm EW}\le 0.73$, $-25^\circ\le\phi\le 25^\circ$, 
and $-0.04\le\varepsilon_a\cos\eta\sin^2\!\gamma\le 0$. 
Figure~\ref{fig:CKMfit} shows the allowed regions in the 
$(\bar\rho,\bar\eta)$ plane for the representative values $X_R=0.25$, 
0.75, and 1.25 (from right to left). We stress that with this method 
a useful constraint on the Wolfenstein parameters is obtained for 
any value of $X_R$. 

\begin{figure}[htb]
   \epsfxsize=7cm
   \centerline{\epsffile{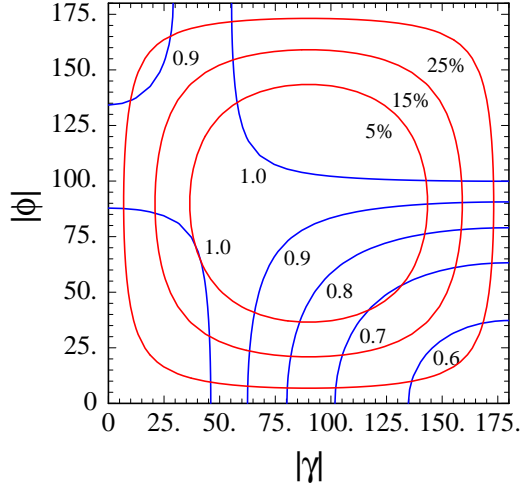}}
\caption{\label{fig:contours}
Contours of constant $R_*$ (``hyperbolas'') and constant 
$|\widetilde A|$ (``circles'') in the $(|\gamma|,|\phi|)$ plane. 
The sign of the asymmetry $\widetilde A$ determines the sign of the 
product $\sin\gamma\sin\phi$. The contours for $R_*$ refer to values 
from 0.6 to 1.0 in steps of 0.1, those for the asymmetry correspond 
to 5\%, 15\%, and 25\%, as indicated.}
\end{figure}

\vspace{0.2cm}\noindent
{\em Model-independent determination:}\\
It is important that, once more precise data on $B^\pm\to\pi K$ 
decays will become available, it will be possible to test the
prediction of a small strong phase $\phi$ experimentally. To this 
end, one must determine the CP asymmetry $\widetilde A$ defined in 
(\ref{Atil}) in addition to the ratio $R_*$. From (\ref{expr}) it 
follows that for fixed values of $\bar\varepsilon_{3/2}$ and 
$\delta_{\rm EW}$ the quantities $R_*$ and $\widetilde A$ define 
contours in the $(\gamma,\phi)$ plane, whose intersections determine 
the two phases up to possible discrete ambiguities~\cite{us2,me}. 
Figure~\ref{fig:contours} shows these contours for some 
representative values, assuming $\bar\varepsilon_{3/2}=0.21$, 
$\delta_{\rm EW}=0.64$, and $\varepsilon_a=0$. In practice, including 
the uncertainties in the values of these parameters changes the 
contour lines into contour bands. Typically, the spread of the bands 
induces an error in the determination of $\gamma$ of about~\cite{me} 
$10^\circ$. In the most general case there are up to eight discrete 
solutions for the two phases, four of which are related to the other 
four by a sign change $(\gamma,\phi)\to(-\gamma,-\phi)$. However, for 
typical values of $R_*$ it turns out that often only four solutions 
exist, two of which are related to the other two by a sign change. 
The theoretical prediction that $\phi$ is small implies that solutions 
should exist where the contours intersect close to the lower portion 
in the plot. Other solutions with large $\phi$ are strongly 
disfavored. Note that according to (\ref{expr}) the sign of the CP 
asymmetry $\widetilde A$ fixes the relative sign between the two 
phases $\gamma$ and $\phi$. If we trust the theoretical prediction 
that $\phi$ is negative~\cite{newfact}, it follows that in most cases 
there remains only a unique solution for $\gamma$, i.e.\ the 
CP-violating phase $\gamma$ can be determined without any discrete 
ambiguity. 

Consider, as an example, the hypothetical case where $R_*=0.8$ and 
$\widetilde A=-15\%$. Figure~\ref{fig:contours} then allows the four 
solutions where $(\gamma,\phi)\approx(\pm 82^\circ,\mp 21^\circ)$ or 
$(\pm 158^\circ,\mp 78^\circ)$. The second pair of solutions is 
strongly disfavored because of the large values of the strong phase 
$\phi$. From the first pair of solutions, the one with 
$\phi\approx-21^\circ$ is closest to our theoretical expectation that 
$\phi\approx -11^\circ$, hence leaving $\gamma\approx 82^\circ$ as 
the unique solution.

\section{Sensitivity to New Physics}
\label{sec:6}

In the presence of New Physics the theoretical description of 
$B^\pm\to\pi K$ decays becomes more complicated. In particular, new 
CP-violating contributions to the decay amplitudes may be induced. 
A detailed analysis of such effects has been presented in~\cite{troja}. 
A convenient and completely general parameterization of 
the two amplitudes in (\ref{ampls}) is obtained by replacing
\begin{equation}\label{replace}
   P \to P' \,, \qquad
   \varepsilon_a\,e^{i\gamma} e^{i\eta} \to
    i\rho\,e^{i\phi_\rho} \,, \qquad
   \delta_{\rm EW} \to a\,e^{i\phi_a} + ib\,e^{i\phi_b} \,,
\end{equation}
where $\rho$, $a$, $b$ are real hadronic parameters, and $\phi_\rho$, 
$\phi_a$, $\phi_b$ are strong phases. The terms $i\rho$ and $ib$ 
change sign under a CP transformation. New Physics effects 
parameterized by $P'$ and $\rho$ are isospin conserving, while those 
described by $a$ and $b$ violate isospin symmetry. Note that the 
parameter $P'$ cancels in all ratios of branching ratios and thus does 
not affect the quantities $R_*$ and $X_R$ as well as any CP 
asymmetry. Because the ratio $R_*$ in (\ref{Rst}) would be 1 in the 
limit of isospin symmetry, it is particularly sensitive to 
isospin-violating New Physics contributions. 

New Physics can affect the bound on $\gamma$ derived from 
(\ref{Rbound}) as well as the extraction of $\gamma$ using the 
strategies discussed above. We will discuss these two possibilities 
in turn.

\subsection{Effects on the Bound on $\gamma$}

The upper bound on $R_*^{-1}$ in (\ref{Rbound}) and the 
corresponding bound on $X_R$ shown in Fig.~\ref{fig:Rbound} are 
model-independent results valid in the Standard Model. Note that 
the extremal value of $R_*^{-1}$ is such that 
$|X_R|\le(1+\delta_{\rm EW})$ irrespective of $\gamma$. A value of 
$|X_R|$ exceeding this bound would be a clear signal for New 
Physics~\cite{me,Mati,troja}. 

Consider first the case where New Physics may induce arbitrary 
CP-violating contributions to the $B\to\pi K$ decay amplitudes, 
while preserving isospin symmetry. Then the only change with respect 
to the Standard Model is that the parameter $\rho$ may no longer be 
as small as $O(\varepsilon_a)$. Varying the strong phases $\phi$ and 
$\phi_\rho$ independently, and allowing for an arbitrarily 
large New Physics contribution to $\rho$, one can derive the 
bound~\cite{troja} 
\begin{equation}\label{phiarb}
   |X_R| \le \sqrt{1 - 2\delta_{\rm EW}\cos\gamma 
   + \delta_{\rm EW}^2} \le 1+\delta_{\rm EW} \,.
\end{equation}
The extremal value is the same as in the Standard Model, 
i.e.\ isospin-conserving New Physics effects cannot lead to a value 
of $|X_R|$ exceeding $(1+\delta_{\rm EW})$. For intermediate values 
of $\gamma$ the Standard Model bound on $X_R$ is weakened; but even 
for large $\rho=O(1)$, corresponding to a significant New 
Physics contribution to the decay amplitudes, the effect is small.

If both isospin-violating and isospin-conser\-ving New Physics 
contributions are present and involve new CP-violating phases, the 
analysis becomes more complicated. Still, it is possible to derive 
model-independent bounds on $X_R$. Allowing for arbitrary values 
of $\rho$ and all strong phases, one obtains~\cite{troja}
\begin{eqnarray}\label{abbound}
   |X_R| &\le& \sqrt{(|a|+|\cos\gamma|)^2 + (|b|+|\sin\gamma|)^2}
    \nonumber\\
   &\le& 1 + \sqrt{a^2 + b^2}
    \le \frac{2}{\bar\varepsilon_{3/2}} + X_R \,,
\end{eqnarray}
where the last inequality is relevant only in cases where
$\sqrt{a^2 + b^2}\gg 1$. The important point to note is that with
isospin-violating New Physics contributions the value of $|X_R|$
can exceed the upper bound in the Standard Model by a potentially 
large amount. For instance, if $\sqrt{a^2+b^2}$ is twice as large
as in the Standard Model, corresponding to a New Physics contribution 
to the decay amplitudes of only 10--15\%, then $|X_R|$ could be as 
large as 2.6 as compared with the maximal value 1.8 allowed (for 
arbitrary $\gamma$) in the Standard Model. Also, in the most general 
case where $b$ and $\rho$ are non-zero, the maximal value $|X_R|$ can 
take is no longer restricted to occur at the endpoints 
$\gamma=0^\circ$ or $180^\circ$, which are disfavored by the global 
analysis of the unitarity triangle~\cite{BaBar}. Rather, $|X_R|$ 
would take its maximal value if $|\tan\gamma|=|\rho|=|b/a|$.

The present experimental value of $X_R$ in (\ref{Rval}) has too 
large an error to determine whether there is any deviation from the 
Standard Model. If $X_R$ turns out to be larger than 1 (i.e.\ at
least one third of a standard deviation above its current central 
value), then an interpretation of this result in the Standard Model 
would require a large value $|\gamma|>91^\circ$ (see 
Fig.~\ref{fig:Rbound}), which would be difficult to accommodate in 
view of the upper bound implied by the experimental constraint on 
$B_s$--$\bar B_s$ mixing, thus providing evidence for New Physics. 
If $X_R>1.3$, one could go a step further and conclude that the New 
Physics must necessarily violate isospin~\cite{troja}. 

\subsection{Effects on the Determination of $\gamma$}

A value of the observable $R_*$ violating the bound (\ref{Rbound}) 
would be an exciting hint for New Physics. However, even if a future 
precise measurement will give a value that is consistent with the 
Standard Model bound, $B^\pm\to\pi K$ decays provide an excellent 
testing ground for physics beyond the Standard Model. This is so 
because New Physics may cause a significant shift in the value of 
$\gamma$ extracted using the strategies discussed earlier, leading 
to inconsistencies when this value is compared with other 
determinations of $\gamma$. 

A global fit of the unitarity triangle combining information from 
semi-leptonic $B$ decays, $B$--$\bar B$ mixing, CP violation in the 
kaon system, and mixing-induced CP violation in $B\to J/\psi\,K_S$ 
decays provides information on $\gamma$ which in a few years will 
determine its value within a rather narrow range~\cite{BaBar}. Such 
an indirect determination could be complemented by direct 
measurements of $\gamma$ using, e.g., $B\to D K^{(*)}$ decays, or 
using the triangle relation $\gamma=180^\circ-\alpha-\beta$ 
combined with a measurement of $\alpha$. We will assume that a 
discrepancy of more than $25^\circ$ between the ``true'' 
$\gamma=\mbox{arg}(V_{ub}^*)$ and the value $\gamma_{\pi K}$ 
extracted in $B^\pm\to\pi K$ decays will be observable after a few 
years of operation at the $B$ factories. This sets the benchmark for 
sensitivity to New Physics effects.

\begin{figure}[htb]
   \epsfxsize=8cm
   \centerline{\epsffile{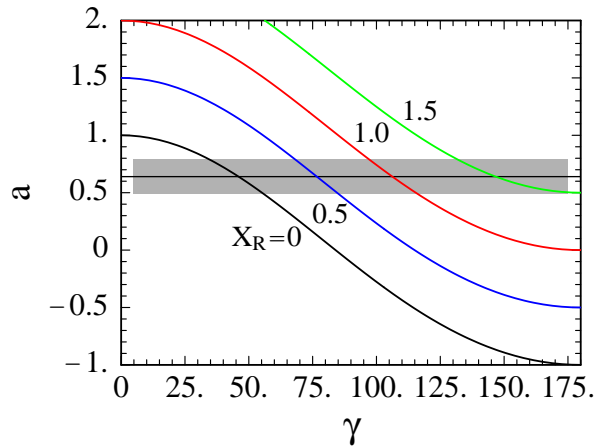}}
\caption{\label{fig:shift}
Contours of constant $X_R$ versus $\gamma$ and the parameter $a$, 
assuming $\gamma>0$. The horizontal band shows the value of $a$ 
in the Standard Model.}
\end{figure}

In order to illustrate how big an effect New Physics could have 
on the extracted value of $\gamma$, we consider the simplest case 
where there are no new CP-violating couplings. Then all New Physics 
contributions in (\ref{replace}) are parameterized by the single 
parameter $a_{\rm NP}\equiv a-\delta_{\rm EW}$. A more general 
discussion can be found in Ref.~127.
%\citelow{troja} 
We also assume for 
simplicity that the strong phase $\phi$ is small, as suggested by 
(\ref{phiest}). In this case the difference between the value 
$\gamma_{\pi K}$ extracted from $B^\pm\to\pi K$ decays and the 
``true'' value of $\gamma$ is to a good approximation given by
\begin{equation}
   \cos\gamma_{\pi K} \simeq \cos\gamma - a_{\rm NP} \,.
\end{equation}
In Fig.~\ref{fig:shift} we show contours of constant $X_R$ versus 
$\gamma$ and $a$, assuming without loss of generality that 
$\gamma>0$. Obviously, even a moderate New Physics contribution to 
the parameter $a$ can induce a large shift in $\gamma$. Note that 
the present central value of $X_R\approx 0.7$ is such that values of 
$a$ less than the Standard Model result $a\approx 0.64$ are 
disfavored, since they would require values of $\gamma$ exceeding 
$100^\circ$, in conflict with the global analysis of the unitarity 
triangle~\cite{BaBar}.

\subsection{Survey of New Physics models}

In Ref.~127,
%\citelow{troja} 
we have explored how New Physics could affect purely hadronic FCNC 
transitions of the type $\bar b\to\bar s q\bar q$ focusing, in 
particular, on isospin violation. Unlike in the Standard Model, where 
isospin-violating effects in these processes are suppressed by 
electroweak gauge couplings or small CKM matrix elements, in many New 
Physics scenarios these effects are not parametrically suppressed 
relative to isospin-conserving FCNC processes. In the language of 
effective weak Hamiltonians this implies that the Wilson coefficients 
of QCD and electroweak penguin operators are of a similar magnitude. 
For a large class of New Physics models we found that the coefficients 
of the electroweak penguin operators are, in fact, due to ``trojan'' 
penguins, which are neither related to penguin diagrams nor of 
electroweak origin. 

Specifically, we have considered: (a) models with tree-level FCNC 
couplings of the $Z$ boson, extended gauge models with an extra $Z'$ 
boson, supersymmetric models with broken R-parity; (b) supersymmetric 
models with R-parity conservation; (c) two-Higgs-doublet models, and 
models with anomalous gauge-boson couplings. Some of these models 
have also been investigated in Refs.~125 and 126.
%\citelow{CDK,anom} 
In case (a), the 
electroweak penguin coefficients can be much larger than in 
the Standard Model because they are due to tree-level processes. In 
case (b), these coefficients can compete with the ones of the 
Standard Model because they arise from strong-interaction box 
diagrams, which scale relative to the Standard Model like
$(\alpha_s/\alpha)(m_W^2/m_{\rm SUSY}^2)$. In models (c), on 
the other hand, isospin-violating New Physics effects are not
parametrically enhanced and are generally smaller than in the 
Standard Model.

\begin{table}
\caption{\label{tab:1}
Maximal contributions to $a_{\rm NP}$ and shifts in $\gamma$ in 
extensions of the Standard Model. For the case of supersymmetric 
models with R-parity the first (second) row corresponds to maximal 
right-handed (left-handed) strange-bottom squark mixing. For the 
two-Higgs-doublet models we take $m_{H^+}>100$\,GeV and 
$\tan\beta>1$.}
\vspace{0.5cm}
\centerline{\begin{tabular}{|l|cc|}\hline\hline
Model \rule[-0.25cm]{0cm}{0.7cm} & $|a_{\rm NP}|$
 & $|\gamma_{\pi K}-\gamma|$ \\
\hline
FCNC $Z$ exchange \rule{0cm}{0.4cm} & 2.0 & $180^\circ$ \\
extra $Z'$ boson & 14 & $180^\circ$ \\
SUSY without R-parity & 14 & $180^\circ$ \\
\hline
SUSY with R-parity \rule{0cm}{0.4cm} & 0.4 & $25^\circ$ \\
 & 1.3 & $180^\circ$ \\
\hline
two-Higgs-doublet models \rule{0cm}{0.4cm} & 0.15 & $10^\circ$ \\
anomalous gauge-boson couplings & 0.3 & $20^\circ$ \\[0.06cm]
\hline\hline
\end{tabular}}
\end{table}

For each New Physics model we have explored which region of 
parameter space can be probed by the $B^\pm\to\pi K$ observables, 
and how big a departure from the Standard Model predictions one 
can expect under realistic circumstances, taking into account all 
constraints on the model parameters implied by other processes. 
Table~\ref{tab:1} summarizes our estimates of the maximal 
isospin-violating contributions to the decay amplitudes, as 
parameterized by $|a_{\rm NP}|$. They are the potentially most 
important source of New Physics effects in $B^\pm\to\pi K$ 
decays. For comparison, we recall that in the Standard Model 
$a\approx 0.64$. Also shown are the corresponding maximal values 
of the difference $|\gamma_{\pi K}-\gamma|$. As noted above, in 
models with tree-level FCNC couplings New Physics effects can be 
dramatic, whereas in supersymmetric models with R-parity 
conservation isospin-violating loop effects can be competitive with 
the Standard Model. In the case of supersymmetric models with 
R-parity violation the bound (\ref{abbound}) implies interesting 
limits on certain combinations of the trilinear couplings 
$\lambda_{ijk}'$ and $\lambda_{ijk}''$, as discussed in Ref.~127.
%\citelow{troja}

\section{Concluding Remarks}

We have presented an introduction to recent developments in the
theory and phenomenology of $B$ physics, focusing on heavy-quark 
symmetry, exclusive and inclusive weak decays of $B$ mesons, and 
rare $B$ decays that are sensitive to CP-violating weak phases of
the Standard Model. The theoretical tools that allow us to perform 
quantitative calculations are various forms of heavy-quark expansions, 
i.e.\ expansions in logarithms and inverse powers of the large scale
provided by the heavy-quark mass, $m_b\gg\Lambda_{\rm QCD}$. 

Heavy-flavor physics is a rich and diverse area of current research, 
which is characterized by a fruitful interplay between theory and 
experiments. This has led to many significant discoveries and 
developments. $B$ physics has the potential to determine many 
important parameters of the electroweak theory and to test the 
Standard Model at low energies. At the same time, through the study
of CP violation it provides a window to physics beyond the Standard
Model. Indeed, there is a fair chance that such New Physics will
first be seen at the $B$ factories, before it can be explored in 
future collider experiments at the Tevatron and the Large Hadron 
Collider.

\vspace{0.3cm}
{\em Acknowledgements:\/}
It is a great pleasure to thank the Organizers of the Trieste Summer 
School in Particle Physics for the invitation to present these 
lectures and for providing a stimulating and relaxing atmosphere, 
which helped to initiate many physics discussions. In particular, I
wish to express my gratitude to Gia Dvali, Antonio Masiero, Goran 
Senjanovic and Alexei Smirnov for their great hospitality and their 
many efforts to make my stay in Trieste a memorable one. Last but not 
least, I wish to thank the students of the school for their lively 
interest in these lectures.

\end{document}